\documentclass[iop,usenatbib,useAMS,numberedappendix,tighten,appendixfloats]{emulateapj}
\bibpunct{(}{)}{;}{a}{}{,}
\usepackage{apjfonts}
\usepackage{enumerate,amsmath}

\usepackage[T1]{fontenc}
\usepackage[latin1]{inputenc}
\usepackage{graphicx}
\usepackage{amssymb}
\usepackage{times}

\newcommand{\angstrom}{\text{\normalfont\AA}}
\newcommand{\tauv}{\hbox{$\hat{\tau}_{V}$}}

\newcommand{\mic}{\hbox{$\mu$m}}
\newcommand{\lsun}{\hbox{$L_\odot$}}
\newcommand{\msun}{\hbox{$M_\odot$}}

\newcommand{\magphys}{\hbox{\sc magphys}}
\newcommand{\hyperz}{\hbox{\sc hyper-z}}

\def\lesssim{\mathrel{\hbox{\rlap{\hbox{\lower5pt\hbox{$\sim$}}}\hbox{$<$}}}}
\def\gtrsim{\mathrel{\hbox{\rlap{\hbox{\lower5pt\hbox{$\sim$}}}\hbox{$>$}}}}
 
 \DeclareMathSymbol{\la}{3}{AMSa}{46}
 \DeclareMathSymbol{\ga}{3}{AMSa}{38}

\shorttitle{The physical properties of ALESS SMGs}
\shortauthors{E. da Cunha et al.}

\begin{document}


\title{An ALMA survey of Sub-millimeter Galaxies in the Extended Chandra Deep Field South: Physical properties derived from ultraviolet-to-radio modelling}


\author{E. da Cunha\altaffilmark{1,2},
F. Walter\altaffilmark{1}, I. R. Smail\altaffilmark{3}, A. M. Swinbank\altaffilmark{3}, J. M. Simpson\altaffilmark{3},
R. Decarli\altaffilmark{1}, J. A. Hodge\altaffilmark{4}, A. Weiss\altaffilmark{5}, \\
P. P. van der Werf\altaffilmark{6},
F. Bertoldi\altaffilmark{7}, S. C. Chapman\altaffilmark{8}, P. Cox\altaffilmark{9}, A. L. R. Danielson\altaffilmark{3},
H. Dannerbauer\altaffilmark{10}, T. R. Greve\altaffilmark{11}, \\
R. J. Ivison\altaffilmark{12,13}, A. Karim\altaffilmark{7}, A. Thomson\altaffilmark{3}}
\affil{$^{1}$Max-Planck-Institut f\"ur Astronomie, K\"onigstuhl 17, 69117 Heidelberg, Germany}
\affil{$^{2}$Centre for Astrophysics and Supercomputing, Swinburne University of Technology, Hawthorn, Victoria 3122, Australia}
\affil{$^{3}$Institute for Computational Cosmology, Durham University,  South Road,  Durham DH1 3LE,  United Kingdom}
\affil{$^{4}$National Radio Astronomy Observatory, 520 Edgemont Rd., Charlottesville VA, USA}
\affil{$^{5}$Max-Planck-Institut f\"ur Radioastronomie, Auf dem H\"ugel 69, 53121 Bonn, Germany}
\affil{$^{6}$Leiden Observatory, Leiden University, P.O. Box 9513, 2300 RA Leiden, The Netherlands}
\affil{$^{7}$Argelander Institute for Astronomy, University of Bonn, Auf dem H\"ugel 71, 53121 Bonn, Germany}
\affil{$^{8}$Department of Physics and Atmospheric Science, Dalhousie University, Halifax, NS B3H 3J5, Canada}
\affil{$^{9}$Atacama Large Millimeter/Submillimeter Array, Alonso de Cordova 3107, Vitacura, Santiago, Chile}
\affil{$^{10}$Universit\"at Wien, Institut f\"ur Astrophysik, T\"urkenschanzstra{\ss}e 17, A-1180 Wien, Austria}
\affil{$^{11}$Department of Physics and Astronomy, University College London, Gower Street, London WC1E 6BT, UK}
\affil{$^{12}$European Southern Observatory, Karl-Schwarzschild Stra{\ss}e 2, D-85748 Garching bei M\"unchen, Germany}
\affil{$^{13}$ Institute for Astronomy, University of Edinburgh, Blackford Hill, Edinburgh EH9 3HJ, United Kingdom}

\email{Electronic address: edacunha@swin.edu.au}

\begin{abstract}
\noindent

The ALESS survey has followed-up a sample of 122 sub-millimeter sources in the Extended {\em Chandra} Deep Field South at 870\mic\ with ALMA, allowing to pinpoint the positions of sub-millimeter galaxies (SMGs) to $\sim0.3$~arcsec and to find their precise counterparts at different wavelengths. This enabled the first compilation of the multi-wavelength spectral energy distributions (SEDs) of a statistically reliable survey of SMGs. In this paper, we present a new calibration of the \magphys\ SED modelling code that is optimized to fit these ultraviolet-to-radio SEDs of $z>1$ star-forming galaxies using an energy balance technique to connect the emission from stellar populations, dust attenuation and dust emission in a physically consistent way. We derive statistically and physically robust estimates of the photometric redshifts and physical parameters (such as stellar masses, dust attenuation, star formation rates, dust masses) for the ALESS SMGs. We find that the ALESS SMGs have a median stellar mass $M_\ast = (8.9\pm0.1) \times10^{10}~M_\odot$, median star formation rate SFR $= 280\pm70~M_\odot\,\mathrm{yr}^{-1}$, median overall $V$-band dust attenuation $A_V=1.9\pm0.2$~mag, median dust mass $M_\mathrm{dust}=(5.6\pm1.0)\times10^8~M_\odot$, and median average dust temperature $T_\mathrm{dust}\simeq40$~K. We find that the average intrinsic spectral energy distribution of the ALESS SMGs resembles that of local ultra-luminous infrared galaxies in the infrared range, but the stellar emission of our average SMG is brighter and bluer, indicating lower dust attenuation, possibly because they are more extended. We explore how the average SEDs vary with different parameters (redshift, sub-millimeter flux, dust attenuation and total infrared luminosity), and we provide a new set of SMG templates that can be used to interpret other SMG observations.
To put the ALESS SMGs into context, we compare their stellar masses and star formation rates with those of less actively star-forming galaxies at the same redshifts. We find that, at $z\simeq2$, about half of the SMGs lie above the star-forming main sequence (with star formation rates three times larger than normal galaxies of the same stellar mass), while half are consistent with being at the high-mass end of the main sequence. At higher redshifts ($z\simeq3.5$), the SMGs tend to have higher star formation rates and stellar masses, but the fraction of SMGs that lie significantly above the main sequence decreases to less than a third.

\end{abstract}


\section{Introduction} \label{sect:intro}

The first sensitive sub-millimeter bolometer camera (the Sub-millimeter Common User Bolometer Array, SCUBA) unveiled a new population of galaxies over a decade ago \citep{Smail1997,Hughes1998, Barger1998}, with very large sub-millimeter fluxes ($>1$~mJy at 850\mic), which were named sub-millimeter galaxies (SMGs)\footnote{In this paper we will refer to our sources as SMGs using this purely observational definition i.e. our sources are SMGs because they were selected to be bright in the sub-millimeter band. In this context, this definition does not carry any further assumptions on the nature and intrinsic physical properties of these sources.}. These galaxies were later identified to be typically at high redshift, and often very faint or completely undetected at (rest-frame) optical wavelengths (e.g.~\citealt{Smail2002,Dannerbauer2002,Walter2012,Simpson2013}). The large sub-millimeter fluxes of SMGs imply very large dust infrared luminosities ($L_\mathrm{dust} > 10^{12}~\lsun$), which are likely powered by intense star formation (exceeding several times $100~\msun \mathrm{yr}^{-1}$). Even though the number density of these galaxies is low (e.g.~\citealt{Weiss2009}), such highly star-forming galaxies are thought to be the progenitors of local elliptical galaxies (e.g.~\citealt{Smail2002,Simpson2013}), and the intense gas consumption in these objects may be linked to AGN growth and feedback (e.g.~\citealt{Hopkins2008}). Therefore, understanding these objects is crucial to trace the evolution of todays massive galaxies (see \citealt{Blain2002,Casey2014} for reviews).

A main limitation in characterizing sub-millimeter galaxies in terms of redshift and physical properties such as stellar mass, star formation rate and dust attenuation has been identifying their counterparts at shorter wavelengths, due to the large beams of the single-dish sub-millimeter discovery observations. Modern (sub-)millimeter interferometers such as the IRAM Plateau de Bure Interferometer (PdBI) and now the Atacama Large Millimeter Array (ALMA) are allowing us to pin-point the position of SMGs with unprecedented accuracy, which allows, for the first time, to reliably identify their counterparts at optical, infrared and radio wavelengths.

The single-dish LABOCA 870-\mic\ survey of the Extended {\em Chandra} Deep Field South (ECDF-S), LESS, is the largest and deepest contiguous survey ever performed at that wavelength, and identified 126 sub-millimeter sources with 870-\mic\ fluxes above 4~mJy \citep{Weiss2009}. 122 of the 126 LESS sources were followed-up at the same wavelength, with unprecedentedly high sensitivity and spatial resolution enabled by ALMA in the Cycle 0 program ALESS \citep{Karim2013,Hodge2013}. The high resolution enabled by the ALMA interferometer allowed for a de-blending of multiple sources that were previously identified as a single source due to the large beam of single-dish observations, and to pin-point the location of the detected SMGs to within $\sim0.3$~arcsec \citep{Hodge2013}. Using these observations, \cite{Hodge2013} demonstrated the limitations in identifying the counterparts of sub-millimeter sources in previous single-dish studies. The ALESS observations showed that between 35 and 50\% of the detected LABOCA sources are actually multiple sources blended in the large beam, and that 45\% of counterparts were missed by previous statistical methods that relied on higher-resolution 24-\mic\ and radio observations. {\em These observations make ALESS the first statistically reliable survey of SMGs, which allows for a complete and unbiased multi-wavelength study of the properties of this galaxy population.}

The ECDF-S field contains a wealth of ancillary data from X-rays, to optical, near-infrared, far-infrared and radio wavelengths. \cite{Simpson2013} measured aperture photometry of the ALESS counterparts (based on the ALMA positions) in up to 19 bands from the $U$-band to the {\it Spitzer}/IRAC 8-\mic\ band. They used the observed optical spectral energy distributions (SEDs) to compute photometric redshifts, stellar masses and dust attenuation of the sources. At longer wavelengths, \cite{Swinbank2013} estimated the infrared fluxes of the ALESS SMGs by using the ALMA positions as priors to de-blend the lower-resolution {\em Herschel} photometry, thus building the full infrared SEDs from 24~\mic\ to 500-\mic, also including the ALMA flux at 870\mic\ and the radio flux at 1.4~GHz. \cite{Swinbank2013} fit these SEDs using templates to derive the first estimates of total infrared luminosity, dust temperature and star formation rates of these galaxies.

The ALESS sample and the studies described above allow us to build, for the first time, the complete spectral energy distributions from (rest-frame) ultraviolet to the radio of a complete, statistically unbiased survey of SMGs with robust counterparts. However, so far most studies have analysed the (optical) stellar and the infrared (dust) SEDs of SMGs independently from each other (with the exception \citealt{Michalowski2010}, who used GRASIL \citep{Silva1998} templates to model the full SEDs of SMGs). Typically, photometric redshifts, stellar masses and dust attenuations are derived from optical data, while dust luminosities, dust masses and star formation rates are constrained from far-infrared/sub-millimeter data. Moreover, often the models used to analyze the optical SEDs are models that were calibrated mostly using `normal' local galaxies which are not likely to be as dust-obscured and actively star-forming as high-redshift SMGs. This is problematic because stellar age, dust and photometric redshift suffer degeneracies (as pointed out by, e.g.~\citealt{Dunlop2007}), and at this high optical depth regime stellar mass-to-light ratios are very difficult to constrain (as discussed by e.g.~\citealt{Hainline2011,Simpson2013}).

In this paper, we interpret consistently the full ultraviolet-to-radio spectral energy distributions of the ALESS SMGs in terms of photometric redshift, stellar content, star formation activity and dust properties using an updated version of the \magphys\ model \citep{daCunha2008}. The new version of the code extends the SED parameter priors to the high-redshift, high-optical depth and actively star-forming regime, mainly by including new star formation histories, a more general prior for the dust attenuation parameters, and the effect of intergalactic medium absorption of ultraviolet photons at high redshift, allowing us to make a detail exploration of the parameter space for our sources. Additionally, our new infrared emission models include a wider range of possible dust temperatures and a simple radio emission component. The new SED-fitting code used in this paper also allows for the photometric redshift to be left as a free parameter when fitting the observed SEDs. This allows us to explore how well we can constrain the physical properties and photometric redshift of SMGs when using all available information from the ultraviolet to the radio.
This paper is organized as follows. In \S\ref{data}, we describe the multi-wavelength data available for the ALESS SMGs. In \S\ref{modelling}, we describe the \magphys\ model, which computes the stellar and dust emission of galaxies in a consistent way, and the modifications to the model that were made to better explore the parameter space of high-redshift galaxies. In \S\ref{sedfits}, we present our method to fit the multi-wavelength SEDs of the ALESS sources using \magphys, discuss the photometric redshifts obtained using our approach, and discuss the properties of the most optically-faint ALESS sources. In \S\ref{properties}, we analyze the physical properties related to the stellar content and dust properties of our sources obtained with \magphys. We discuss the average intrinsic spectral energy distributions of the ALESS SMGs, and how the ALESS SMGs compare with `main sequence' galaxies at their redshifts in \S\ref{discussion}. Our conclusions are summarized in \S\ref{conclusion}.

Throughout this paper, we use a concordance $\Lambda$CDM cosmology with $H_0=70$~km~s$^{-1}$~Mpc$^{-1}$, $\Omega_\Lambda=0.7$ and $\Omega_m=0.3$. Unless otherwise stated, we adopt a \cite{Chabrier2003} initial mass function (IMF) and AB system magnitudes.

\section{Multi-wavelength data}
\label{data}

Of the 122 LESS sources that were followed-up at higher resolution with ALMA as part of the ALESS program, several were resolved into multiple counterparts, producing a total of 131 SMG detections \citep{Hodge2013}. 99 of these ALESS SMGs are considered to be the most reliable by \cite{Hodge2013}, based on the fact that they are significant detections (with S/N$>3.5$) inside the ALMA primary beam of good-quality maps -- these are the sources in the {\sc main} ALESS catalog. The remaining 32 sources ({\sc supplementary} catalog) are found to be less reliable by \cite{Hodge2013}, and are therefore are not included in the {\sc main} ALESS sample (\citealt{Karim2013} show that up to 30\% of SMGs in the {\sc supplementary} catalog are likely to be spurious).

We choose to include only the 99 SMGs from the {\sc main} ALESS catalog in our analysis, because they are the most statistically reliable sample of SMGs and therefore allow for a complete and unbiased study of this population of galaxies.
The positions of these SMGs are known to $\sim0.3$~arcsec thanks to the high-resolution ALMA interferometric observations, which allows us to identify the counterparts and measure the fluxes of the SMGs at other wavelengths, using deep ancillary data in the optical, infrared and radio.

\subsection{Optical and near-infrared}
\label{optobs}

To sample the rest-frame stellar emission of our galaxies, we use the (aperture) photometry of the ALESS counterparts (based on the ALMA positions) compiled by \cite{Simpson2013} from archival ground-based and {\em Spitzer}/IRAC observations of the ECDF-S; this provides photometry in up to 19 bands for each galaxy. Most of the $UBVRIzJHK$ photometry comes from the MUSYC survey \citep{Taylor2009}, which is supplemented with (deeper) $U$-band data from the GOODS/VIMOS imaging survey \citep{Nonino2009}. In the near-infrared, additional deep $J$- and $Ks$-band data from the ESO-VLT/HAWK-I survey (Zibetti et al., in prep.) and the CFHT/WIRCAM Taiwan ECDFS NIR Survey (TENIS; \citealt{Hsieh2012}).
The {\em Spitzer}/IRAC data at 3.6, 4.5, 5.0 and 8.0~\mic\ comes from the {\em Spitzer}/IRAC MUSYC Public Legacy in the ECDFS survey (SIMPLE; ~\citealt{Damen2011}). From these images, seeing- and aperture-matched photometry was measured by \cite{Simpson2013} across all filters by centering 3~arcsec apertures on the ALMA position (with a possible shift of $<0.5$~arcsec) and applying aperture corrections based on the total flux of a composite PSF. The uncertainties are computed by measuring the flux in 3-arcsec apertures placed in random blank regions of the sky in each image (and assuming that sky noise dominates the flux uncertainty); full details are given in \cite{Simpson2013}.

77 of our ALESS main catalog sources are detected (i.e. their flux is 3$\sigma$ above the background noise) in at least four optical/near-IR bands\footnote{We refer to these SMGs as the `optically-bright sample' and to the remaining 22 SMGs which are detected in fewer than four optical/near-IR bands as the `optically-faint sample'.}. 
\citealt{Simpson2013} apply systematic offsets to the measured photometry to optimize SED fits to the \hyperz\ code templates \citep{Bolzonella2000}. Since these offsets are SED-model/fitting-procedure dependent, we chose to use the raw photometry measured in \cite{Simpson2013} without including the offsets (in practice this means that we subtract the offsets from the photometry quoted in their Table 2). In order to account for possible zero point offsets and/or mis-matches between the photometry at different wavelengths, we add a magnitude error in quadrature to the quoted errors that is proportional to the photometric offsets determined by \cite{Simpson2013}. For the photometric bands where \cite{Simpson2013} apply smaller offets, we conservatively add an additional $0.10$~mag to the error in quadrature.

\subsection{Mid-infrared, far-infrared and radio}

We use the mid-infrared, far-infrared and radio fluxes of the ALESS sources compiled in \cite{Swinbank2013}, which we briefly summarize here. 

\cite{Swinbank2013} exploit the {\it Spitzer}/MIPS 24-\mic\ images of the ECDFS publicly available from the Far-Infrared Deep Extragalactic (FIDEL) survey, and extracted a catalog of about 3600 sources down to a $5\sigma$ depth of $56\mu$Jy. The Very Large Array (VLA) radio observations at 1.4~GHz come from catalogue of the ECDFS described in \cite{Miller2013}.
The ECDF-S was imaged with {\em Herschel} at 70, 100 and 160\mic\ with PACS as part of the PACS Evolutionary Probe (PEP) survey \citep{Lutz2011}. \cite{Swinbank2013} obtained the PACS fluxes for the ALESS SMGs by matching them with the PEP catalog extracted by \cite{Magnelli2013}. {\em Herschel}/SPIRE imaging at 250, 350 and 500\mic\ is available from the {\em Herschel} Multitiered Extragalactic Survey (HerMES; \citealt{Oliver2012}). Due to the large beam of the SPIRE observations, it is challenging to measure the photometry of individual sources. \cite{Swinbank2013} obtained the SPIRE fluxes of the ALESS sources using a de-blending method based on positional priors from the ALMA maps together with the MIPS 24-\mic\ and radio source catalogs.

We note that, since the resolution of the optical/near-infrared ($\sim1 - 2$~arcsec) and far-infrared ($\sim15 - 25$~arcsec) are so different, it is not possible to measure the photometry in an identical manner across all wavelengths. However, in all cases, we extract `total fluxes': in the optical/near-infrared, \cite{Simpson2013} achieve this by applying aperture corrections based on the total flux of a composite PSF determined using point sources in the images, as mentioned in Section~\ref{optobs}, and described in detail in section~2.2.1 of \cite{Simpson2013}. In the far-infrared, \cite{Swinbank2013} de-blend the {\em Herschel} maps by using a PSF at each wavelength for each galaxy in the prior catalog, and they demonstrate (using simulations) that the total fluxes of sources are recovered using this method. Thus, we expect the combination of far-infrared and optical/near-infrared here to be as reliable as possible given the available data.

\section{The SED model}
\label{modelling}

\begin{figure*}
\begin{center}
\includegraphics[width=\textwidth]{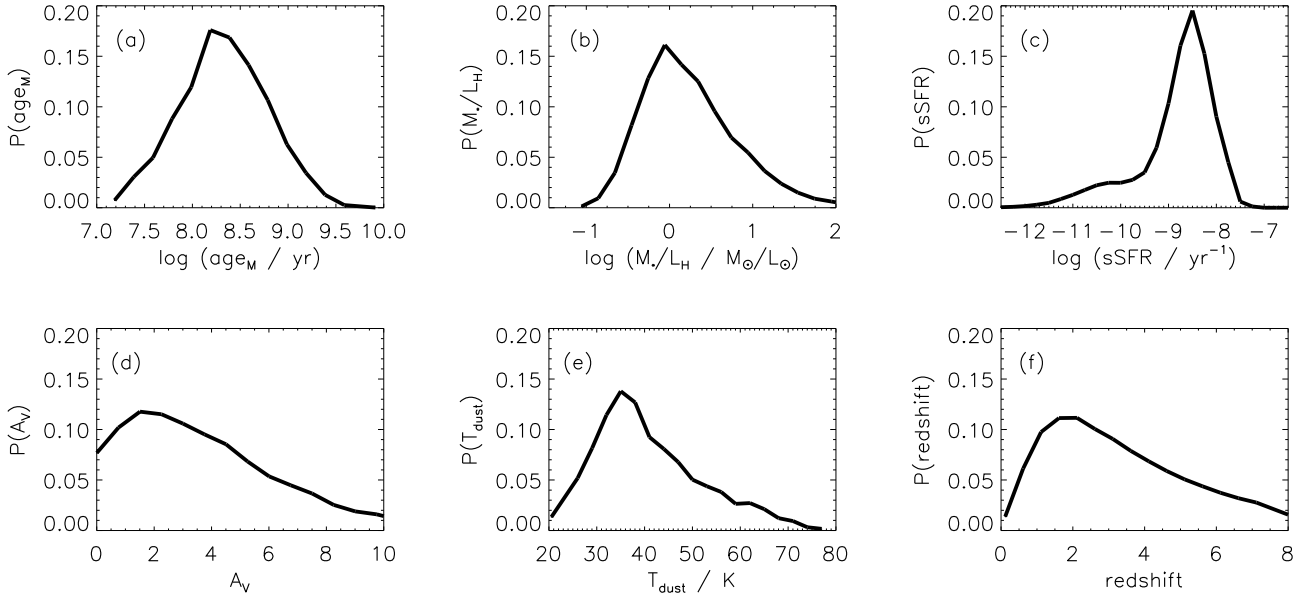}
\vspace{-4.5cm}
\caption{Distribution of some key physical properties in our stochastic library of models: (a) mass-weighted age (defined as in Eq.~\ref{age_m}); (b) mass-to-light ratio in the $H$-band; (c) specific star formation rate; (d) overall $V$-band attenuation (defined as in Eq.~\ref{a_v}); (e) luminosity-averaged dust temperature (defined as in Eq.~\ref{tdust}); and (f) redshift. We note that the shape of distributions of all properties shown here (except redshift) results from the priors that we chose for other parameters of the model that are used to compute them and therefore it is natural that they are not flat (see more details on the chosen priors and how the parameters are computed in \S\ref{modelling}).}
\label{priors}
\end{center}
\end{figure*}

In this paper we aim to model the full observed spectral energy distributions of the ALESS SMGs from the rest-frame ultraviolet to the radio in a physically consistent way. To do so, we use the \magphys\ code\footnote{Publicly available at www.iap.fr/magphys.} \citep{daCunha2008}, which relies on an energy balance technique to consistently combine the emission by stellar populations with the attenuation and emission by dust in galaxies. In this section, we summarize the main ingredients of \magphys\ and present an updated version, which extends the parameter space of the models in order to include also properties that are more likely applicable to high-redshift SMGs. Specifically, we extend our star formation history and dust optical depth priors (\S\ref{sfh} and \S\ref{attenuation}), and we add intergalactic medium absorption in the ultraviolet (\S\ref{igm}).

The stellar population properties of SMGs have been challenging to constrain, mostly because of limited (rest-frame optical/near-infrared) observations available, and also because previous analysis have not modelled the full SEDs in a physically consistent way while sampling properly the full parameter space of star formation histories, metallicities, and dust content. Previous studies that have attempted to derive the stellar masses of high-redshift SMGs by modelling their rest-frame ultraviolet to near-infrared SEDs have used limited parameterizations of the star formation history \citep{Hainline2011,Michalowski2012}. These studies demonstrate that determining the mass-to-light ratio of SMGs is particularly challenging because of the large degeneracy between stellar age and dust reddening, which is exacerbated in the very dust-obscured regime; also, in this regime, a large fraction of the mass may be completely dust-obscured in optically-thick regions.
One of the main goals of our SED modelling is to properly take into account these degeneracies and uncertainties in order to get robust likelihood distributions for the stellar masses of the ALESS SMGs. To do so, we consider a wide range of SFHs, dust attenuations and metallicities, and employ a Bayesian approach to constrain the physical parameters from observations. In addition, by modelling the infrared dust emission consistently with the emission by stellar populations in the rest-frame ultraviolet to near-infrared, we expect to better constrain the attenuation by dust and hence get more insight into obscured populations that are required to power the observed dust emission \citep{daCunha2008,daCunha2010b}.

In the next sub-section, we describe the stellar emission models used to fit the UV to near-IR emission from our galaxies. These models are build using parameter priors that we optimized to get the best possible stellar mass, dust attenuation, mean stellar ages and star formation rate estimates and, at the same time, understand uncertainties and degeneracies that affect these parameters.

\subsection{UV to near-IR: stellar emission}
\label{stellar}

We build a library of $50,000$ stellar emission models that is specifically adapted to interpret the emission from high-redshift, possibly dusty sources with unknown ages and star formation histories. The prior likelihood distributions for each parameter described below are intended to extend the parameter space of the models (compared to the current publicy-available \magphys\ priors which were calibrated on local galaxies of typically lower infrared luminosities and star formation rates; \citealt{daCunha2008,daCunha2010a}), so that they include higher dust optical depths, higher star formation rates, and younger ages (since we expect galaxies to be younger at high redshift). However, it is important to note that we avoid biasing our estimates by also including a large fraction of models extending to low dust optical depths, low star formation rates and older ages, and allowing our Bayesian fitting constrain the most likely values. By doing a Bayesian analysis and building the posterior likelihood distribution of each parameter (based on $\chi^2$ comparison with the observed photometry), we ensure robust estimates and specially robust confidence ranges that take into account parameter degeneracies and uncertainties due to lack of/poor observational constraints in certain wavelength ranges (see more detailed discussion in e.g.~\citealt{daCunha2008}).

We compute the stellar emission of each model as in \cite{daCunha2008}; essentially, the luminosity per unit wavelength emerging at at time $t$ from a model galaxy is expressed as:
\begin{equation}
L_\lambda^\mathrm{\,em}(t)=\int_0^t dt^\prime \Psi(t-t^\prime)\,l^\mathrm{SSP}_\lambda(t^\prime,Z) \,\exp[-\hat\tau_\lambda(t^\prime)] \,,
\label{llambda}
\end{equation}
where $l^\mathrm{SSP}_\lambda(t^\prime,Z)$ is the luminosity emitted per unit wavelength per unit mass by a simple stellar population (SSP) of age $t^\prime$ and metallicity $Z$, $\Psi(t-t^\prime)$ is the star formation rate evolution with time (i.e. the star formation history), and $\hat\tau_\lambda(t^\prime)$ is the effective absorption optical depth seen by stars of age $t^\prime$. 
We compute the emission by SSPs using the spectral population synthesis models of \cite{Bruzual2003} (using a \citealt{Chabrier2003} initial mass function); we adopt a uniform prior in metallicity from 0.2 to 2 times solar. We discuss the modelling of star formation histories and dust attenuation in the next sections.

\subsubsection{Star formation histories}
\label{sfh}

For each model, we parameterize the star formation history (SFH) as a continuous delayed exponential function of the form:
\begin{equation}
\psi_\mathrm{cont}(t) \propto  \gamma^2 t \exp({-\gamma t})\,,
\label{sfh}
\end{equation}
where $t$ is the model age (i.e. the time since the onset of star formation) and $\gamma=1/\tau_{\mathrm{SF}}$ is the inverse of the star formation time-scale in Gyr$^{-1}$. This form of the star formation history for high-redshift galaxies is motivated by \cite{Lee2010}. There is growing evidence (based on observations and on galaxy evolution models) that, for high redshift galaxies, the SFHs are likely to have a variety of shapes, but overall they should be rising with time instead of declining as the `$\tau$-models' typically used at low redshift \citep{Pacifici2013,Behroozi2013,Simha2014}. The SFH parameterization we adopt here (Eq.~\ref{sfh}) rises linearly at early ages and then declines exponentially with the timescale defined by the $\gamma$ parameter. For each model, we draw $\gamma$ randomly from a broad prior distribution ($\gamma$ can vary between $0.075$ and $1.5$, which corresponds to the peak of the SFH happening between 0.7 and 13.3~Gyrs after the onset of star formation). We draw the age randomly between 0.1 and 10~Gyr.
This ensures that our model library includes a wide range of possible continuous SFHs that go from essentially $\tau$-like models peaking at very early ages, to approximately constant star formation rates with time, to SFHs that are linearly increasing with time. Despite the fact that we cover a wide range of star formation evolution with this prior, the actual form of SFHs is likely to be more complex than any analytical parameterization (see e.g. \citealt{Conroy2013,Simha2014}). To account for stochasticity on the star formation histories, we superimpose star formation bursts of random duration and amplitude to the continuous model defined in Eq.~\ref{sfh}. We set the probability of a burst of star formation occurring at any random time in the previous 2~Gyrs to 75\%; each burst can last between 30 and 300~Myr, and the total mass of stars formed during the burst can have any random value between 0.1 and 100 times the mass formed by the underlying continuous SFH (Eq.~\ref{sfh}). This ensures that we account for as wide a range of SFHs as possible using our simple parameterization (including both starburst-like and more quiescently star-forming), which is crucial to sample all possible stellar ages and mass-to-light ratios in our analysis, and thus get the most robust constraints on these parameters for the observed galaxies.

In this context, the `age' of a galaxy model is simply the time when the star formation history starts in our analytical parameterization and has no real physical meaning. In order to have a more reliable measure of the overall age of the stellar population in our galaxies, we define the `mass-weighted age' of each model as:
\begin{equation}
\mathrm{age}_M=\frac{\int_0^t dt^\prime \,t^\prime \Psi(t-t^\prime)}{\int_0^t dt^\prime \Psi(t-t^\prime)} \,,
\label{age_m}
\end{equation}
where $\Psi(t-t^\prime)$ is the star formation history of each model, which is essentially $\psi_\mathrm{cont}(t-t^\prime)$ (Eq.~\ref{sfh}) plus random bursts.
The value of $\mathrm{age}_M$ hence depends not only on the model age, but also strongly on the shape of the star formation history. In Fig.~\ref{priors}(a), we plot the distribution of $\mathrm{age}_M$ in our model library, that results from the different random SFHs and model ages included. We note that $R$-band light-weighted ages (i.e. the ages of the stars dominating the rest-frame $R$-band light) for the models in our library are typically lower than the mass-weighted ages, with $\mathrm{age}_R/\mathrm{age}_M=0.82\pm0.30$.

We define the current star formation rate (SFR) of each model as the average of the star formation history $\Psi$ over the last 10~Myr.

\subsubsection{Dust attenuation}
\label{attenuation}

We use the two-component model of \cite{Charlot2000} to describe the attenuation of stellar emission at ultraviolet, optical and near-infrared wavelengths. This model accounts for the fact that young stars that are still inside their birth clouds are more dust-attenuated than intermediate-age and old stars in the diffuse ISM. In practice, the effective optical depth in Eq.~\ref{llambda} is described as:
\begin{equation}
\hat{\tau}_\lambda(t^\prime)=
\begin{cases}
\hat{\tau}_\lambda^{\,\mathrm{BC}}+\hat{\tau}_\lambda^{\,\mathrm{ISM}} \, & \text{for $t^\prime\leq t_\mathrm{BC}$,} \\
\hat{\tau}_\lambda^{\,\mathrm{ISM}}\, & \text{for $t^\prime> t_\mathrm{BC}$}\,,
\end{cases}
\end{equation}
where
\begin{equation}
\hat{\tau}_\lambda^{\,\mathrm{BC}}=(1-\mu)\,\hat\tau_V\,(\lambda/5500\angstrom)^{-1.3}
\end{equation}
is the effective attenuation optical depth of dust in stellar birth clouds, and
\begin{equation}
\hat{\tau}_\lambda^{\,\mathrm{ISM}}=\mu\,\hat\tau_V\,(\lambda/5500\angstrom)^{-0.7}
\end{equation}
is the effective attenuation optical depth in the diffuse ISM. In practice, the three free parameters of our dust attenuation model are the lifetime of stellar birth clouds $t_\mathrm{BC}$, the effective (i.e.~angle-averaged) $V$-band optical depth seen by stars younger than $t_\mathrm{BC}$ in the birth clouds \tauv, and $\mu$, the fraction of \tauv\ seen by stars older than $t_\mathrm{BC}$. 
We change the priors of these parameters compared to previous studies \citep{daCunha2008,daCunha2010a} in order to reflect our uncertainty of stars/dust geometry and optical depth in high-redshift SMGs and to ensure that we cover a broad parameter space. We allow for dust attenuation to reach higher values both in the stellar birth clouds and the diffuse ISM, to account for the fact that SMGs might be heavily optically-thick (such as local ULIRGs; \citealt{daCunha2010b}). The prior probability distribution for \tauv\ is constant between 0 and 6, and declines exponentially to a maximum of 20; for $\mu$, we set the prior probability distribution to be a Gaussian centered at 0.25 with 0.10 standard deviation. We let the lifetime of birth clouds $t_\mathrm{BC}$ vary between 5 and 50 Myr -- typically this was fixed at 10~Myr \citep{Charlot2000}; since this is an unknown parameter and the ISM conditions in SMGs might be different than in moderately star-forming local galaxies (e.g. higher densities, stronger stellar winds from the starburst), we vary this unknown parameter so that the spectral fitting solution marginalizes over a wide range of possible values.

Even though our dust attenuation prescription is described by the three parameters described above, for simplicity and to allow for a more direct comparison with other studies, we also define, for each model, the resulting overall $V$-band dust attenuation as
\begin{equation}
A_V=-2.5\log\frac{L_V^\mathrm{\,em}}{L_V^\mathrm{\,int}}\,,
\label{a_v}
\end{equation}
where $L_V^\mathrm{\,obs}$ is the emitted (i.e. observed) $V$-band luminosity, and $L_V^\mathrm{\,int}$ is the intrinsic (i.e. dust-free) $V$-band luminosity of each model. The priors of \tauv, $\mu$ and $t_\mathrm{BC}$ described above result in the prior distribution for $A_V$ plotted in Fig.~\ref{priors}(d).
 
\subsubsection{IGM absorption}
\label{igm}
We include absorption by the intergalactic medium (IGM) in our model SEDs, which strongly affects the rest-frame ultraviolet emission from high-redshift galaxies. To do so, we use the absorption prescription from \cite{Madau1995}, which includes Lyman series line blanketing and Lyman-continuum absorption. To account for different opacities along different lines of sight (which reflect different distributions and properties of absorbers along the line of sight), we draw the IGM effective absorption optical depth of each model $\tau_\mathrm{eff}^\mathrm{\,IGM}(z)$ from a Gaussian distribution centered at the mean IGM effective absorption absorption given in \cite{Madau1995}, $\tau_\mathrm{eff}^\mathrm{\,IGM,Madau}(z)$, with a standard deviation of $0.5$.

\subsection{Dust emission and radio component}
\label{dust}

We model the emission by dust from rest-frame mid-infrared to millimeter wavelengths using four main dust components as described in \cite{daCunha2008}: (i) a polycyclic aromatic hydrocarbons (PAHs) empirical template; (ii) mid-infrared continuum from hot dust; (iii) warm dust in thermal equilibrium; and (iv) cold dust in thermal equilibrium. The dust components in thermal equilibrium emit as modified black bodies [$\nu^\beta B_\nu(T)$], with emissivity index $\beta$ fixed at 1.5 for the warm components and 2.0 for the cold components. The equilibrium of warm dust in stellar birth clouds $T_w^\mathrm{\,BC}$ is uniformly distributed in the range $30-80$~K, and the equilibrium temperature of cold dust in the `diffuse ISM' $T_c^\mathrm{\,ISM}$ is uniformly distributed in the range $20-40$~K. The relative contributions of these dust emission components to the total infrared luminosity are also free parameters of the model.

We note that the detailed PAH emission in SMGs may be different from our adopted template, and there may even be variations of the PAH emission between SMGs, since the strength of different features depends on factors such as the intensity of the radiation field and metallicity (e.g.~\citealt{Draine2001,Draine2007}). Previous observations of the mid-infrared emission from $z\sim2$ SMGs by \cite{Delmestre2009} show that their average mid-infrared spectrum is similar to that of local starbursts, which implies that our PAH template is at least a good approximation. \cite{Delmestre2009} also find some variation of the ratios of different PAH features from SMG to SMG that we are not able to reproduce using our fixed template, however due to the lack of data sampling the (rest-frame) mid-infrared emission for our SMGs in detail, we would not be able to constrain that part of the spectrum, even if we included variation of the PAH emission in our modelling.
Furthermore, while the shape of the PAH emission in our model is fixed, the contribution of PAHs to the total dust luminosity is allowed to vary, therefore the models include a range of overall PAH emission strength. While this parameter itself can be hard to constrain in the absence of photometric data sampling the rest-frame mid-infrared, thanks to the MIPS 24-\mic\ data (combined with the {\em Herschel} and ALMA data which constrain the total infrared SED), we are able to loosely constrain the PAH contribution for our galaxies. This contribution is typically less than 10\% of the total infrared luminosity, and therefore uncertainties related to the PAH template should not affect the energy balance significantly.

Even though our infrared SEDs contain multiple temperature components, for simplicity, and to make comparisons between galaxies easier, we define an average, {\em luminosity-weighted dust temperature} for each model as:
\begin{equation}
T_\mathrm{dust}= \frac{\xi_w^\mathrm{\,tot} \times T_\mathrm{w}^\mathrm{\,BC} + \xi_c^\mathrm{\,tot} \times T_c^\mathrm{\,ISM} + \xi_w^\mathrm{\,ISM} \times T_w^\mathrm{\,ISM} \times f_\mu}{\xi_w^\mathrm{\,tot}+ \xi_c^\mathrm{\,tot} + \xi_w^\mathrm{\,ISM} \times f_\mu} \,,
\label{tdust}
\end{equation}
where $\xi_w^\mathrm{\,tot}$ and $\xi_c^\mathrm{\,tot}$ are the fraction of total dust luminosity contributed by the (birth cloud) warm and (diffuse ISM) cold dust components, respectively, and $f_\mu$ is the fraction of total dust luminosity contributed by the diffuse ISM component (which are all free parameters of the model). $\xi_w^\mathrm{\,ISM}$ and $T_w^\mathrm{\,ISM}$ are the fractional contribution and temperature of warm dust to the cool ISM component, and are fixed at $\xi_w^\mathrm{\,ISM}=0.07$ and $T_w^\mathrm{\,ISM}=45$~K (see model details in \citealt{daCunha2008}). By varying the temperatures and relative contributions of the warm and cold dust to the total dust emission, we obtain the distribution of average dust temperatures in our model library plotted in Fig.~\ref{priors}(e).

To fit the observed radio fluxes of the ALESS SMGs, we also add radio emission to our SEDs by following the simple method described in \cite{daCunha2013a}, which essentially uses the radio/far-infrared correlation and fixed slopes for the thermal and non-thermal radio emission for each model (see also \citealt{Dale2002}). To summarize, we compute the radio emission as the sum of a thermal (free-free emission) component with spectral shape $L_\nu^\mathrm{th} \propto \nu^{-0.1}$ and a non-thermal component with shape $L_\nu^\mathrm{nth} \propto \nu^{-0.8}$. We fix the contribution of the thermal component to the 20-cm radio continuum  at 10\%, and scale the total radio emission of each model to the infrared luminosity using the local radio/far-infrared correlation $q$ i.e. the ratio between the far-infrared and the 1.4~GHz flux density. We draw the value of $q$ randomly for each model from a Gaussian prior distribution centered at $q=2.34$ (the local value; \citealt{Yun2001}), with a $\sigma_q=0.25$, in order to account for possible scatter in the radio/far-infrared correlation (e.g.~\citealt{Ivison2010}). A lack of strong redshift evolution of the radio/far-infrared correlation is supported by several studies (e.g.~\citealt{Ibar2008,Ivison2010,Sargent2010,Huang2014,Thomson2014}), and our fits (e.g.~Fig.~\ref{fit_aless1}) show that this assumption is sufficient to reproduce the observed 1.4~GHz flux of most ALESS sources detected in the radio.

\paragraph{AGN contamination.} We note that the SED models used in this study assume that the dominant source of dust heating is star formation, and do not include AGN emission. This should not have a great impact in our analysis of the full sample for two main reasons. First, the fraction of AGN in our SMG sample is likely low, as shown by \cite{Wang2013}, who find that only 10 of the 91 ALESS SMGs in the area covered by {\em Chandra} are X-ray sources, and 8 of those (ALESS011.1, ALESS017.1, ALESS057.1, ALESS066.1, ALESS070.1, ALESS073.1, ALESS084.1, ALESS114.2) are identified as AGN. Second, most of the parameters recovered by \magphys\ are robust to AGN contamination even when the AGN contributes up to 75\% of the ultraviolet-to-infrared emission, as shown in an independent study by \cite{Hayward2015}. The parameter that suffers most uncertainty in the case of strong AGN contamination is the stellar mass, which may be overestimated by \magphys\ by up to 0.3~dex when the contribution of AGN-heated dust to the near-/mid-infrared is the highest (see \citealt{Hayward2015} for more details).

\subsection{Combined ultraviolet-to-radio model SEDs}
\label{combined}

An important feature of our model is that we combine the stellar emission (attenuated by dust) with the dust emission in a self-consistent way using a simple energy balance argument: that the energy absorbed by dust in stellar birth clouds and the diffuse ISM ($L_\mathrm{dust}^\mathrm{\,BC}$ and $L_\mathrm{dust}^\mathrm{\,ISM}$, respectively) is re-radiated in the infrared\footnote{As in \cite{daCunha2008}, dust self-absorption is not included. Dust might be optically-thick to its own radiation on some lines of sight, which is not included in the modelling, since this would require a more complete radiative transfer calculation. This should not significantly affect our results, because dust self-absorption affects mostly the rest-frame mid-infrared spectrum, which is not sampled in detail by our data, and the angle-averaged optical depth values are relatively small. Moreover, dust self-absorption should not affect our energy balance to more than a few percent.}.
Different combinations of physical parameters (star formation histories, dust attenuation parameters) can lead to the same absorbed energies $L_\mathrm{dust}^\mathrm{\,BC}$ and $L_\mathrm{dust}^\mathrm{\,ISM}$ in a model galaxy, and these energies can be distributed in the infrared range in different ways depending on the parameters regulating the contribution and temperatures of the different dust components. In practice, we associate each model in our stellar population library (\S\ref{stellar}) with all the models in the infrared spectral library (\S\ref{dust}) that have similar fraction of the total dust luminosity contributed by the diffuse ISM component, $f_\mu=L_\mathrm{dust}^\mathrm{\,ISM}/[L_\mathrm{dust}^\mathrm{\,BC}+L_\mathrm{dust}^\mathrm{\,ISM}]$ (within an error interval $\delta f_\mu=0.15$ which accounts for uncertainties in dust modelling due to geometry, for example).

\section{Multi-wavelength SED fits}
\label{sedfits}

\subsection{Fitting method}
\label{fitting}

\begin{figure*}
\includegraphics[width=0.9\textwidth]{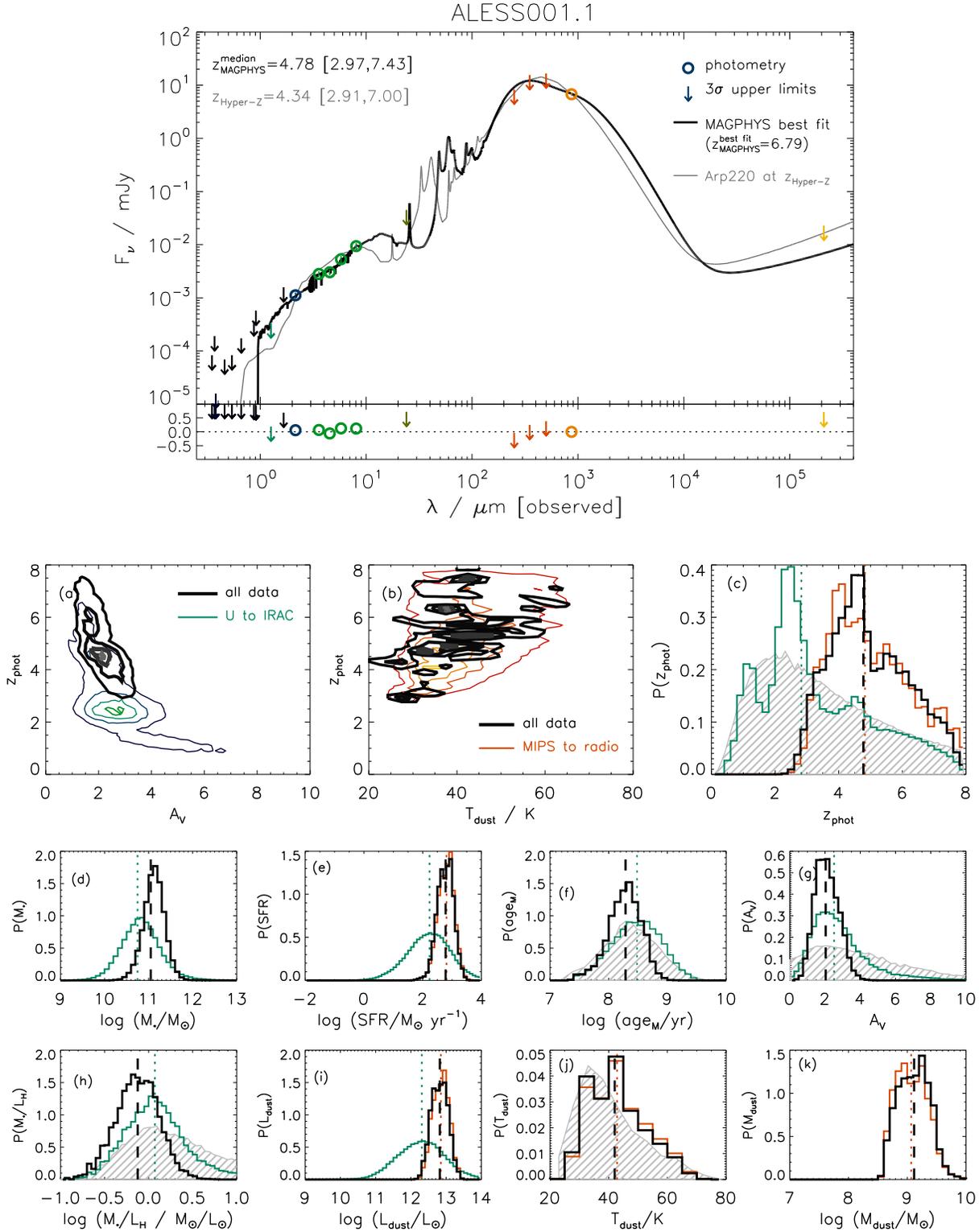}
\caption{Example of the \magphys\ outputs for the galaxy ALESS001.1 (all other SMG fit figures are available on the online version of the journal). {\em Top panel:} Spectral energy distribution fit of the galaxy. The circles show the measured photometry and the arrows indicate $3\sigma$ upper limits, as described in \S\ref{data}. The black line shows our best-fit SED model to all the bands from the ultraviolet to the radio. In the bottom we plot the fit residuals in each band, computed as $(F_\nu^\mathrm{obs}-F_\nu^\mathrm{mod})/F_\nu^\mathrm{obs}$. We also show, for comparison, the SED of the local ULIRG Arp220 scaled to the measured ALMA 870-\mic\ flux at the redshift found by \cite{Simpson2013} by fitting the ultraviolet/optical/near-IR data with \hyperz. In the top-left corner, we indicate our median-likelihood redshift obtained by with \magphys\ (with the 16th--84th percentile range indicated in brackets), and also the \hyperz\ photometric redshift obtained by \cite{Simpson2013} (with their $1\sigma$ range in brackets).
{\em Bottom panels:} (a) joint likelihood distributions of the $V$-band dust attenuation and redshift; (b) joint likelihood distribution of the dust temperature and redshift; the remaining panels show the (normalized) marginalized probability distribution for the redshift (c), stellar mass (d), star formation rate (e), mass-weighted age (f), overall $V$-band dust attenuation (g), $H$-band mass-to-light ratio (h), dust luminosity (i), dust average temperature (j) and dust mass (k). In all these panels, the black lines indicate the probability distributions obtained when fitting the whole ultraviolet-to-radio SED. The green lines and the orange lines show the likelihood distributions we obtain when fitting only the ultraviolet-to-near-infrared (i.e. stellar emission) and mid-infrared-to-radio (i.e. dust emission), respectively. In panels (c), (f), (g), (h) and (j) we also plot, in grey, the prior distribution of each parameter, for comparison.}
\label{fit_aless1}
\end{figure*}

We use the Bayesian method described in \cite{daCunha2008} to compare our library of SED models described in the previous section with the observed photometry in the ultraviolet, optical, infrared and radio for each ALESS source.

The standard version of the \magphys\ code fits the SEDs at fixed redshift. Since spectroscopic redshifts are not available for the whole sample of ALESS SMGs (A. Danielson et al., in prep.), and the uncertainties associated with the optical/near-IR photometric redshifts of \cite{Simpson2013} can be quite large (when few photometric data points are available to constrain the fits), here we leave the redshift as a free parameter in our analysis, and effectively test the use of \magphys, for the first time, as a `photometric redshift code'. This has two advantages that we discuss in detail later: (i) we can incorporate the far-infrared and radio data as additional constraints on the photometric redshift (which is particularly important for the most optically-faint SMGs), and (ii) we can include the uncertainties on the photometric redshift in the error bars of all other physical parameters in a self-consistent way, since the likelihood distributions of the physical properties and redshift are computed simultaneously.

For each model in our library, we compute the predicted flux in our 28 bands (from the $U$-band to 1.4~GHz) for a set of 100 redshifts drawn randomly from the prior distribution plotted in Fig.~\ref{priors}(f), which includes a broad peak at $z\simeq2.5$ (note that if the model age is older than the age of the Universe at a given redshift, than that model+redshift combination is not considered). We use this prior in order to avoid oversampling unlikely regions of the redshift space, however we have checked that using a flat redshift prior does not change our results. At each redshift, the predicted model flux in each band is computed by first applying the IGM absorption prescription to the stellar emission at that redshift as described in \S\ref{igm}, and then convolving the total (stellar+dust emission) model SED in the observed-frame with the filter response functions.
Then, we compare the observed fluxes of our galaxies in all the observed bands with the predicted model fluxes by computing the $\chi^2$ goodness-of-fit for each model in our library (upper limits are included by setting the flux to zero and adopting the upper limit value as the flux uncertainty). We then build the likelihood distribution of each parameter in our model (including the redshift) by marginalizing the probability of each model $P \propto \exp(-\chi^2/2)$ over all other model parameters (more details can be found in \citealt{daCunha2008}). We take our estimates of each parameter to be the median of its likelihood distribution, and the confidence range as the 16th to 84th percentile range.

As an example, in Fig~\ref{fit_aless1}, we plot our best-fit model (i.e. the model that minimizes $\chi^2$) SED obtained with this method, as well as the fit residuals in all bands, for the first source in our catalog, ALESS001.1 (the fit results for all the other sources can be found in the online version of the journal). For comparison, we also plot the SED of the prototypical low-redshift ULIRG, Arp220, normalized at the observed ALMA 870-\mic\ flux, at the photometric redshift of \cite{Simpson2013}, obtained from fitting solely the $U$-band to IRAC fluxes using the \hyperz\ code \citep{Bolzonella2000}.
Assuming that the SMGs have similar SEDs to that of a typical local ULIRG, then if the \hyperz\ photometric redshift is correct, the Arp220 SED should approximately follow the observed photometry in all the other bands. If this is not the case (e.g. ALESS014.1, ALESS015.1 and others) then this means that either the photometric redshift is not accurate, or that the SED of these SMGs is different from that of Arp220 (or both). This is why it is important to fit the whole SED using a wide range of possible SED shapes (which we obtain by varying the physical parameters in our spectral library), and leaving the redshift free, as we do in this study.

In the bottom panels of Fig.~\ref{fit_aless1}, we plot the normalized probability distributions for several parameters (redshift, stellar mass, star formation rate, mass-weighted age, average dust attenuation, $H$-band mass-to-light ratio, total dust luminosity, average dust temperature and dust mass), obtained when fitting the whole SED, and also when fitting only the ultraviolet, optical and near-infrared fluxes, and when fitting only the mid-infrared to radio fluxes. These probability distributions show how tightly our method constrains the redshifts and physical properties of the ALESS SMGs, and also highlight the effect of fitting the whole SED simultaneously versus fitting only the stellar emission or only the infrared emission as done in previous studies of these SMGs (e.g.~\citealt{Simpson2013,Swinbank2013}).
In panels (f), (g), (h) and (j), we also plot the normalized prior distribution of each parameter in our model library (same as shown in Fig.~\ref{priors}), in order to compare directly between the prior and posterior likelihood distribution. The stellar mass, star formation rate, dust luminosity and dust mass depend on the model scaling, and therefore they do not have fixed prior distributions. The stellar mass prior is essentially flat, and the other parameters scale with stellar mass in a way that is determined by the model prior; for example, the SFR will be determined by the stellar mass and by the specific star formation rate prior, as plotted in Fig.~\ref{priors}(c).

Our Bayesian fitting method implies that, if a parameter is unconstrained (because e.g., of a poor sampling of the SED, degeneracies, or large observational uncertainties) the posterior likelihood distribution of a given physical parameter should be the same as the prior distribution. When data (i.e. multi-wavelength broad-band fluxes) are available to constrain a given parameter, the posterior likelihood distribution becomes different, as each model in the prior distribution is now weighted by our well it fits the data, and not all models in the library fit the data equally well. For the majority of our SMGs, the posterior likelihood distributions of the parameters plotted in the bottom panels of Fig.~\ref{fit_aless1} are different from the prior distributions, which shows that these parameters are constrained by the data (with the width of the distributions indicating how well constrained the parameters are, and typically becoming narrower as more data points are available to sample the SED and/or observational uncertainties become smaller). We find that the typically least well constrained parameters in our analysis (which can be seen from the fact that the posterior likelihood distributions often resemble the prior distributions) are the mass-weighted ages and the dust temperatures. It is not surprising that the ages are not tightly constrained given that the rest-frame optical/near-infrared SEDs are often not well sampled by the available data, and that the stellar age is a very challenging parameter to constrain from broad-band data alone (because of age/dust attenuation degeneracies, and the fact that young massive stars outshine older stars). Regarding the dust temperatures, they are not tightly constrained in cases where there are no available {\it Herschel} detections sampling the peak of the dust SED (as is the case for ALESS001.1; Fig.~\ref{fit_aless1}).

\subsection{Photometric redshifts}
\label{redshifts}

\section{Comparison of photometric and spectroscopic redshifts}
\label{zspec_appendix}

\begin{figure}
\centering
\includegraphics[width=0.5\textwidth]{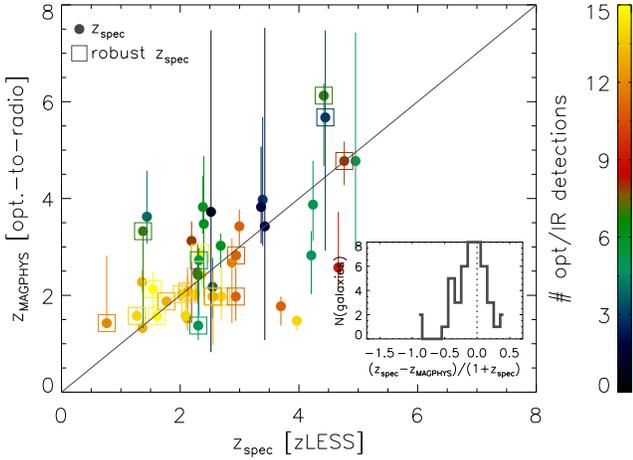}
\caption{Comparison between our median-likelihood redshift estimates and the spectroscopic redshifts available for 45 SMGs in the main ALESS sample from the zLESS program (Danielson et al., in prep.). The 19 sources with the most robust spectroscopic redshifts are highlighted with squares. The vertical error bar shows the 16th -- 84th confidence range of our redshift likelihood distributions. The points are colour-coded by number of available detections in the optical bands (i.e. from $U$ to IRAC-8\mic).}
\label{zspec_comparison}
\end{figure}

\begin{figure}
\centering
\includegraphics[width=0.5\textwidth]{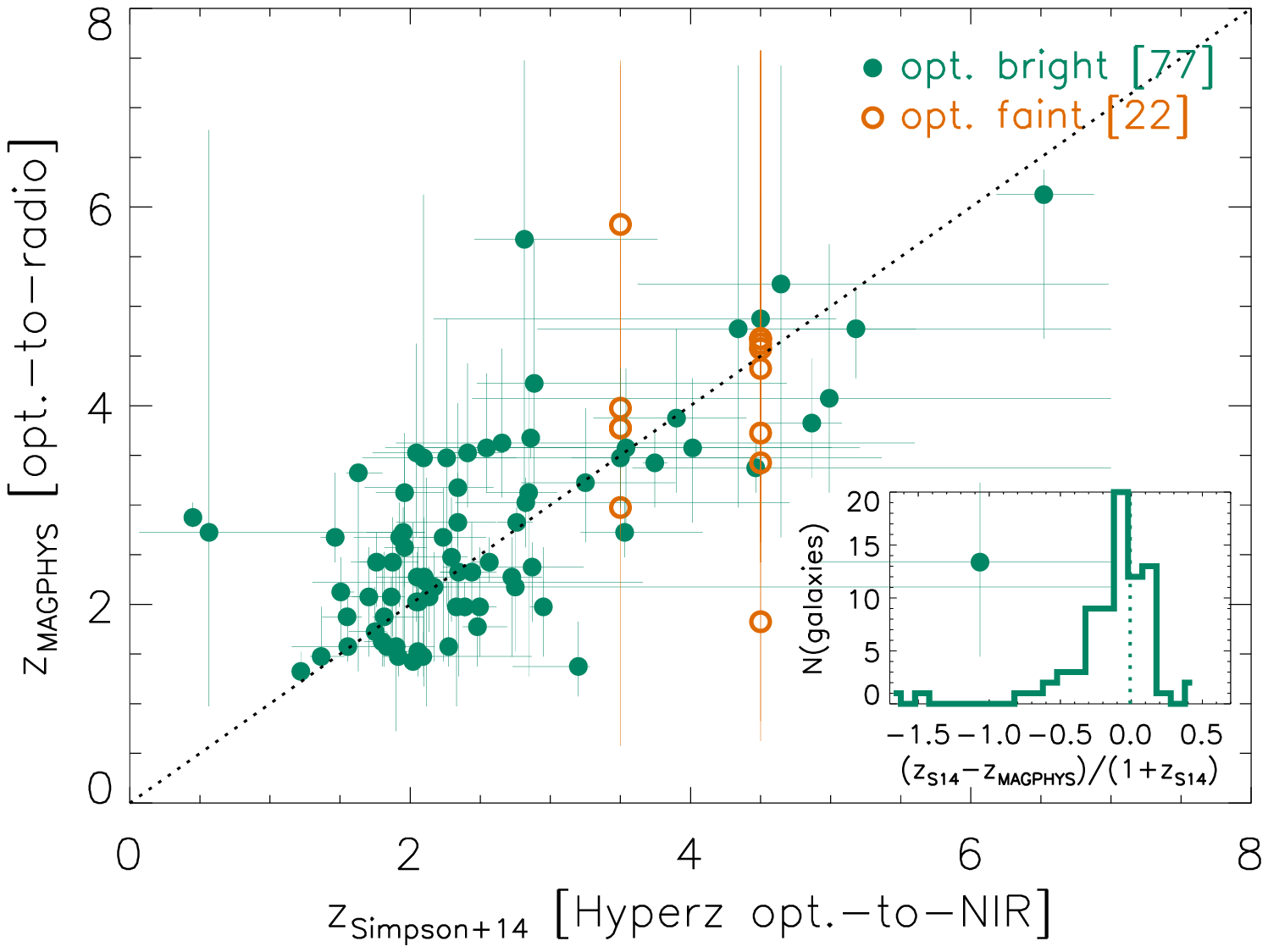}
\caption{Comparison between the \cite{Simpson2013} photometric redshifts obtained by fitting the $U$-band to {\em Spitzer}/IRAC photometry using the \hyperz\ code \cite{Bolzonella2000} ($z_\mathrm{S13}$), and the \magphys\ median-likelihood redshifts obtained from fitting the full $U$-band to radio SEDs ($z_\mathrm{MAGPHYS}$). The error bars show the statistical confidence ranges (for the \magphys\ redshifts, these correspond to the 16th to 84th percentiles of the likelihood distribution marginalized over all other model parameters). The orange symbols correspond to the 22 `optically-faint' SMGs: \cite{Simpson2013} do not constrain individual redshifts for these sources, but they estimate a median redshift of $z_\mathrm{S13}\simeq3.5$ for the SMGs detected in two or three optical bands, and $z_\mathrm{S13}\simeq4.5$ for the SMGs detected in less than two bands.
The inset shows an histogram of the relative difference between the two redshift estimates for the 77 `optically-bright' SMGs for which we have individual estimates from \cite{Simpson2013} (green symbols).
}
\label{zphot_comparison}
\end{figure}

\begin{figure}
\centering
\includegraphics[width=0.5\textwidth]{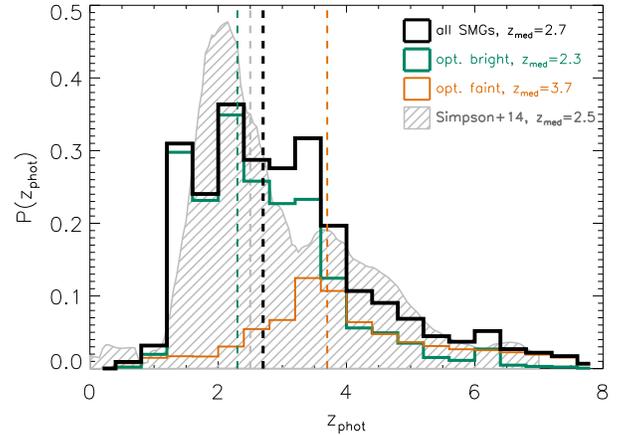}
\vspace{-0.4cm}
\caption{Normalized stacked redshift probability distribution of all 99 SMGs in our sample (in black). The green histogram shows the contribution by the 77 `optically-bright' sources (with four or more detections in the optical/near-infrared range), and the orange histogram shows the contribution by the remaining 22 `optically-faint' sources. In grey, we plot, for comparison, the stacked probability distribution obtained by \cite{Simpson2013} using the \hyperz\ code. The vertical lines show the medians of the distributions (also indicated in the top right corner).}
\label{zstack}
\end{figure}

In the next section we analyze the properties of the ALESS SMGs based on the likelihood distributions of their physical parameters obtained using the method described above. The likelihood distribution of each parameter is obtained by marginalizing the probability over all other parameters, including the redshift. Here we compare our \magphys-based photometric redshift estimates (using the full optical-to-radio SEDs) with the spectroscopic redshifts obtained by A. Danielson et al. (in prep.), and with the photometric redshift estimates in \cite{Simpson2013}, in order to investigate the reliability of our redshift probability distributions before moving on to the analysis of the intrinsic physical properties of the galaxies.

Spectroscopic redshifts are available for 45 of the main sample ALESS galaxies from the zLESS program (A. Danielson et al., in prep.). In Fig.~\ref{zspec_comparison}, we compare our \magphys\ photometric redshift estimates with those spectroscopic redshifts.
We find a generally good agreement between our photometric redshift estimates and the zLESS redshifts. On average, we find a median relative difference of $\Delta z=(z_\mathrm{spec}-z_\mathrm{MAGPHYS})/(1+z_\mathrm{spec})=-0.005$, with a standard deviation of $0.29$. Many of the largest outliers have very large photometric redshift confidence ranges that are still consistent with the spectroscopic redshift. However, there are nine SMGs with $\Delta z > 0.3$ and confidence ranges that do not include the spectroscopic redshifts: ALESS010.1 (robust $z_\mathrm{spec}$), ALESS029.1, ALESS087.1 (robust $z_\mathrm{spec}$), ALESS071.1, ALESS084.1, ALESS069.1, ALESS080.1, ALESS055.1, ALESS037.2.
In the case of ALESS010.1, ALESS084.1, ALESS069.1, the (well-sampled) observed SED seems to be inconsistent with the spectroscopic redshift, particularly the MIPS 24-\mic\ detection tends to constrain the redshift  to be lower. In the case of ALESS037.2, redshift seems inconsistent with the UV/optical upper limits. For ALESS071.1 and ALESS055.1, our \magphys\ fit does not optimally reproduce the IRAC near-infrared data and there is a large observed excess, so probably our fit is not trustworthy. Assuming that the observed SEDs and the spectroscopic redshift are correct, some of these outliers could be cases where the energy balance imposed by \magphys\ does not work properly and so our photometric redshifts may not be reliable. This could happen if the stars/dust geometry in the galaxies is very different than what is assumed in the models, for example if most of the rest-frame UV/optical light comes from a region that is physically separated from where the bulk of the infrared/sub-mm emission originates (e.g.~\citealt{Hodge2012}).

In Fig~\ref{zphot_comparison}, we compare the median-likelihood estimates of the photometric redshift of our sources obtained by fitting the full ultraviolet-to-radio SED with \magphys, $z_\mathrm{MAGPHYS}$, with previous estimates using the \hyperz\ photometric redshift code (which only fits the stellar emission) obtained by \cite{Simpson2013}, $z_\mathrm{S13}$. Overall, we find that the agreement between the redshift estimates is good and there are no strong systematic offsets. The inset histogram in Fig.~\ref{zphot_comparison} shows the distribution of the relative difference between the redshift estimates, which we quantify as $\Delta z=(z_\mathrm{S13}-z_\mathrm{MAGPHYS})/(1+z_\mathrm{S13})$. The median of this difference is only $0.008$ (which is much smaller than the typical error bars), and the standard deviation is $0.32$.
The largest outliers in this comparison are ALESS087.1, ALESS124.1, ALESS061.1, ALESS006.1 (likely gravitationally-lensed), ALESS083.4 (flagged in \citealt{Simpson2013} as possibly having contaminated optical/near-IR photometry), and ALESS065.1 (detected only in three IRAC bands in the optical/near-IR range). In Appendix~\ref{zspec_appendix}, we compare our photometric redshifts with spectroscopic redshifts available for 45 of our SMGs (Danielson et al., in prep).

Our method allows us to investigate in detail the advantage of fitting the full observed $U$-band-to-radio SED simultaneously in deriving the physical properties of galaxies and also photometric redshifts. The bottom panels of Fig.~\ref{fit_aless1} illustrate the differences between parameter likelihood distributions obtained when fitting the full SED and when fitting only the optical-to-near-infrared emission, or only the infrared-to-radio SED. We find that fitting the full SED presents a unique advantage particularly for galaxies for which the number of detections in the optical/near-infrared is low. One example of this is ALESS002.2, which in the optical/near-infrared range is detected only in three IRAC bands. The analysis in \cite{Simpson2013} is unable to constrain the photometric redshift for this source. When we fit only these three optical fluxes using our code, we also obtain a very broad likelihood distribution for the redshift (and also the other physical parameters). The joint redshift-$A_V$ probability distribution shows a clear degeneracy between these two parameters (this is a known degeneracy in photometric redshifts and also noted by \citealt{Dunlop2007}). When we add the infrared data to the fit, since we have four infrared fluxes (from {\em Herschel}/SPIRE and ALMA) constraining the peak of the SED, the redshift is better constrained as well as the other parameters, particularly the dust luminosity and $A_V$ are constrained via energy balance between the optical and infrared.
This is slightly alleviated when more detections sampling the dust emission are available from {\em Spitzer}/MIPS and {\em Herschel}/PACS (e.g.~ALESS014.1, ALESS015.1). When only one infrared flux is available (e.g.~ALESS015.3, ALESS023.7), the redshift is still very hard to constrain, because with only one far-infrared flux we cannot constrain the peak of the SED.

How well each photometric redshift is constrained depends on the number of photometric bands where each galaxy is detected, and also on its intrinsic redshift and SED. Therefore, we characterize the redshift distribution of the whole sample in Fig.~\ref{zstack} by stacking the marginalized redshift likelihood distributions of the galaxies, which naturally accounts for the different redshift uncertainties affecting different sources. The median of this stacked likelihood distribution is $z_\mathrm{phot}=2.7\pm0.1$\footnote{We refer to the \magphys-derived photometric redshifts as $z_\mathrm{phot}$ from now on.}, with a 16th -- 84th percentile range from 1.6 to 4.1. We note that this redshift distribution is consistent with the one derived in \cite{Simpson2013} for the same sample (as shown in Fig.~\ref{zstack}), and also with other SMG samples (e.g.~\citealt{Chapman2005,Aretxaga2007,Wardlow2011,Smolcic2012}).

\subsection{The optically-faint ALESS sources}
\label{section_faint}

\begin{figure*}
\begin{center}
\includegraphics[width=\textwidth]{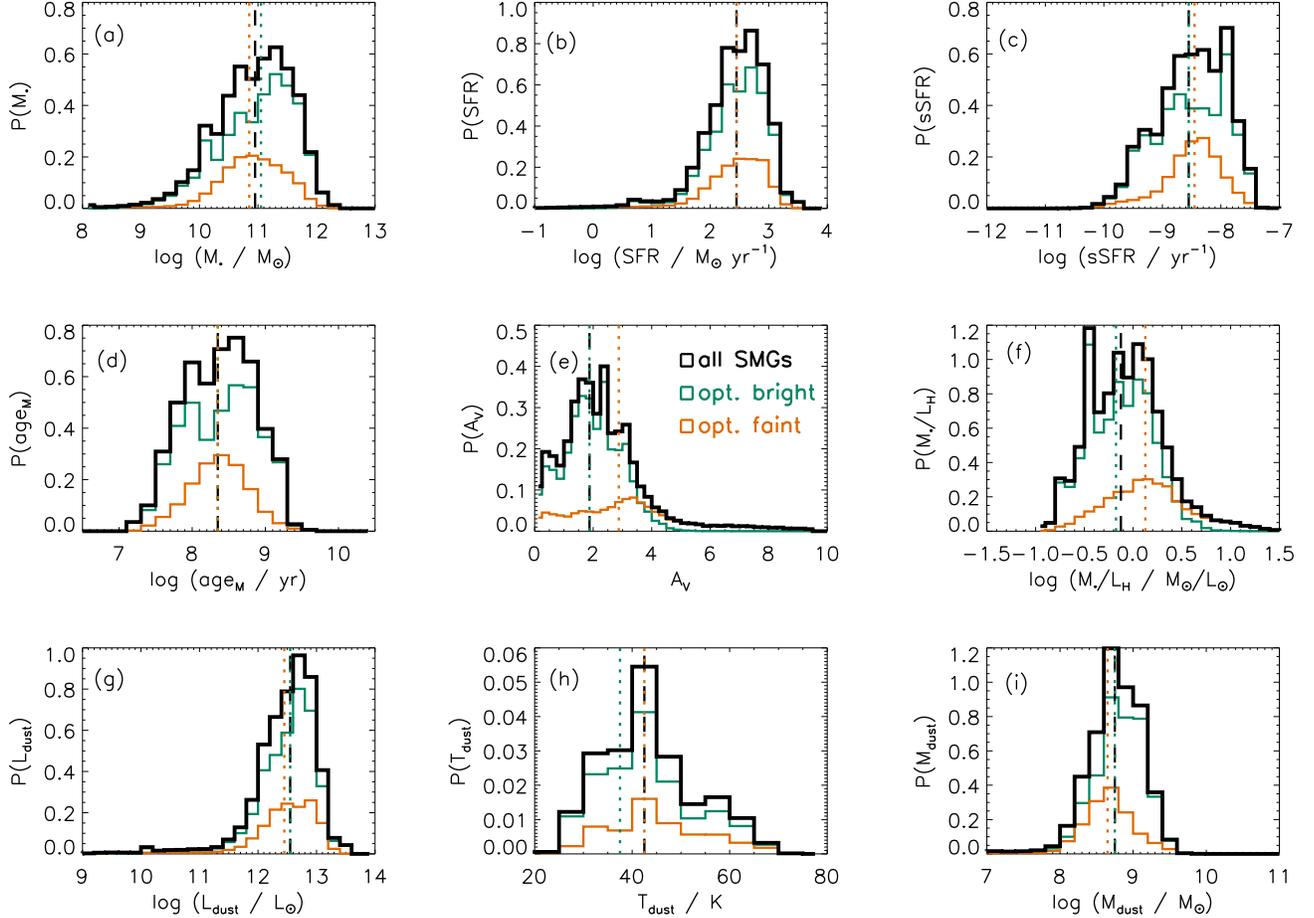}
\caption{Normalized stacked likelihood distributions of different physical parameters for all 99 ALESS SMGs: (a) stellar mass; (b) star formation rate; (c) specific star formation rate; (d) mass-weighted age; (e) average $V$-band dust attenuation; (f) mass-to-light ratio in the $H$ band; (g) total dust luminosity; (h) luminosity-averaged dust temperature; (i) total dust mass. We also plot the contribution by the 77 `optically-bright' SMGs (detected in at least four optical/near-infrared bands; in green) and the 22 `optically-faint' SMGs (detected in fewer than four optical/near-infrared bands; in orange). The vertical lines indicate the median of each distribution (listed also in Table~\ref{parameters_table}). The likelihood distributions show that many of the parameters of the optically-faint sample are unconstrained or consistent with the optically-bright sample. However, the optically-faint sample shows significantly higher average dust attenuation $A_V$ than the optically-bright sample (which also causes the inferred $H$-band mass-to-light ratios to be larger).}
\label{properties_faint}
\end{center}
\end{figure*}

Before we move onto a detailed discussion of the physical properties of all 99 SMGs in the main ALESS catalog (\S\ref{properties}), here we discuss the 22 SMGs that have the lowest detection rate in the optical/near-infrared range, the `optically-faint' sub-sample. These sources are detected in fewer than four bands in the study of the optical/near-infrared emission of the ALESS counterparts of \cite{Simpson2013}. Five of these sources (ALESS035.2, ALESS055.2, ALESS069.3, ALESS087.3, ALES103.3) are undetected in the $U$-band to IRAC-8\mic\ photometry, as well as in the infrared with {\em Spitzer} or {\em Herschel} and in the radio \citep{Swinbank2013}. 

The redshifts and physical properties of these galaxies are particularly challenging to constrain due to the lack of data sampling the SEDs. Our SED modelling approach allows us to attempt to characterize these faint galaxies by combining all the (scarce) available data in the optical/near-infrared and far-infrared/submillimeter, and by including the information available from the upper limits. In Figs.~\ref{zstack} and \ref{properties_faint}, we plot the stacked likelihood distributions thus obtained for these 22 SMGs and compare them to the stacked likelihood distributions of the 77 `optically-bright' sources (i.e. SMGs that are detected in four or more optical/near-infrared bands). We also include the medians and 68\% ranges of these distributions in Table~\ref{parameters_table}.

The likelihood distributions of most properties of the optically-faint SMGs are consistent with them having similar properties to the optically-bright SMGs. The main exceptions are the likelihood distributions of the redshift and average $V$-band dust attenuation $A_V$. The median of the redshift likelihood distribution for the optically-faint SMGs is $z_\mathrm{phot}=3.7\pm0.1$, significantly higher than the median redshift of the optically-bright SMGs ($z_\mathrm{phot}=2.3\pm0.1$). This difference in the average redshift of these sources is consistent with the stacking analysis performed in \cite{Simpson2013}, who concluded that the optically-faint SMGs detected in none or one optical band have an average redshift of 4.5, while the SMGs with two or three detections have an average redshift of 3.5.
The average dust attenuation increases from $A_V=1.9\pm0.3$ for the optically-bright sources to $A_V=2.9\pm0.3$ for the faint sources. These results suggest that the optically-faint ALESS SMGs  may be a subset of the high-redshift ($z>3$) SMG population that have higher dust optical depths. Since our analysis suggests that the total dust mass and total dust luminosity of the sources is similar to the SMGs in the optically-bright sample, then these higher dust optical depths are likely caused not because there is more dust in these galaxies, but because the stars/dust geometry is different. For example, a higher $A_V$ can be a result of edge-on viewing angle, or of a more compact galaxy where dust column is higher; this could be tested by comparing high spatial resolution sub-millimeter continuum imaging of the optically-bright and optically-faint sources at similar redshifts. Our results imply that we could be biasing the redshift distribution of SMGs low when the redshifts are based on detected optical counterparts, because we could be missing this population of very obscured, high-redshift sources that is a non-negligible fraction of all the SMGs in the ALESS sample (22\%). Obtaining redshifts through the detection of molecular lines and/or [CII] in the sub-millimeter is the only reliable way to get the real redshift distribution of these sources (e.g.~\citealt{Walter2012,Weiss2013}).

In the following sections, we analyze the physical properties of the complete sample of 99 ALESS SMGs as a whole, i.e. including both the optically-faint and optically-bright galaxies.

\section{The physical properties of ALESS SMGs}
\label{properties}

\begin{figure*}
\begin{center}
\includegraphics[width=\textwidth]{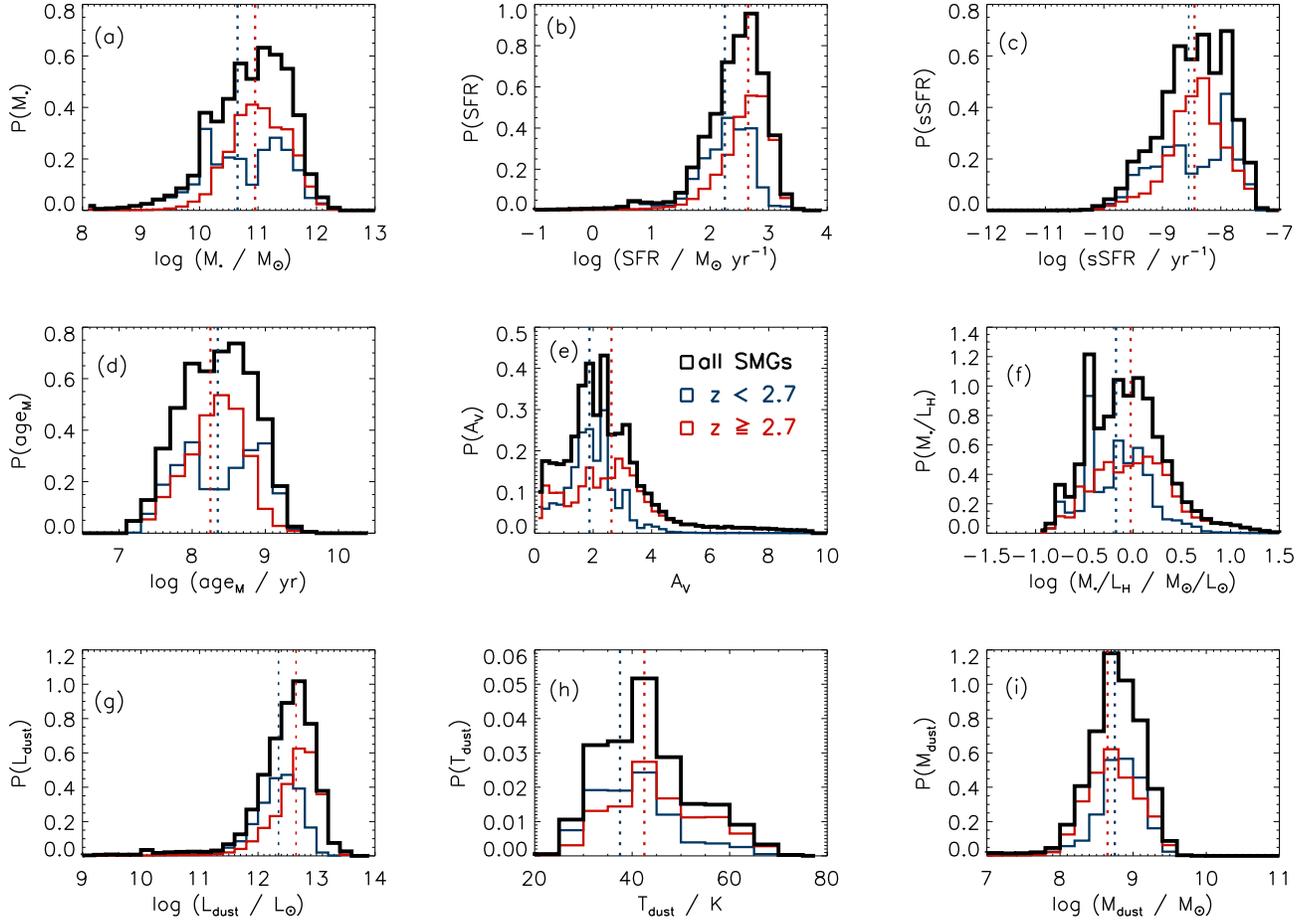}
\caption{Black histograms: normalized stacked likelihood distributions of different physical parameters for the 99 galaxies in our sample, as shown in Fig.~\ref{properties_faint}. The blue histograms show the stacked histograms for the sub-sample of galaxies with $z_\mathrm{phot} < 2.7$ (45 sources), and the red histograms correspond to the sub-sample of galaxies with $z_\mathrm{phot} \ge 2.7$ (54 sources). The medians and 16th--84th percentile ranges of these likelihood distributions are listed on Table~\ref{parameters_table}.}
\label{properties_zcut}
\end{center}
\end{figure*}

\begin{deluxetable*}{lccccc}
\tablecolumns{5}
\tablewidth{0.95\textwidth}
\tablecaption{Average properties of the ALESS SMGs \label{parameters_table}}
\tablehead{
\colhead{} &
\colhead{} &
\colhead{{\sc Whole Sample}} &
\colhead{} &
\colhead{{\sc Optically Bright}} &
\colhead{{\sc Optically Faint} }\\
\\
\colhead{{\sc Parameter}} &
\colhead{all redshifts} &
\colhead{$z_\mathrm{phot} < 2.7$} &
\colhead{$z_\mathrm{phot} \ge 2.7$ } &
\colhead{all redshifts} &
\colhead{all redshifts }\\
\\
\colhead{} &
\colhead{[99 SMGs]} &
\colhead{[45 SMGs]} &
\colhead{[54 SMGs]} &
\colhead{[77 SMGs]} &
\colhead{[22 SMGs]}\\
}
\startdata

$\log\,(\,M_\ast / M_\odot)$ & $10.95^{+0.6}_{-0.8}$  & $10.75^{+0.8}_{-0.8}$ & $10.95^{+0.6}_{-0.4}$ & $11.05^{+0.5}_{-0.9}$ & $10.85^{+0.5}_{-0.6}$\\[0.2cm]

$\log\,(\,\mathrm{SFR} / M_\odot~\mathrm{yr}^{-1})$ & $2.45_{-0.5}^{+0.4}$  & $2.25_{-0.5}^{+0.3}$ & $2.65_{-0.5}^{+0.3}$ & $2.45_{-0.5}^{+0.4}$ & $2.45_{-0.5}^{+0.4}$\\[0.2cm]

$\log\,(\,\mathrm{sSFR} / \mathrm{yr}^{-1})$ & $-8.55_{-0.6}^{+0.6}$  & $-8.65_{-0.7}^{+0.7}$ & $-8.45_{-0.5}^{+0.5}$ & $-8.55_{-0.7}^{+0.6}$ & $-8.45_{-0.5}^{+0.4}$  \\[0.2cm]

$\log\,(\,\mathrm{age}_\mathrm{M} / \mathrm{yr})$ & $8.35_{-0.6}^{+0.5}$ &  $8.45_{-0.7}^{+0.5}$ & $8.35_{-0.5}^{+0.3}$ & $8.35_{-0.6}^{+0.5}$ & $8.35_{-0.4}^{+0.3}$\\[0.2cm]

$ A_V $ & $1.9_{-1.0}^{+1.2}$  &  $1.6_{-0.8}^{+0.8}$ & $2.4_{-1.5}^{+1.5}$ & $1.9_{-1.0}^{+1.0}$ & $2.9_{-1.8}^{+2.2}$\\[0.2cm]

$\log\,(\,M_\ast/L_\mathrm{H}  / M_\odot/L_\odot)$ & $-0.13_{-0.4}^{+0.4}$ &  $-0.18_{-0.3}^{+0.3}$ & $-0.03_{-0.4}^{+0.4}$ & $-0.18_{-0.4}^{+0.4}$ & $0.13_{-0.4}^{+0.4}$\\[0.2cm]

$\log\,(\,L_\mathrm{dust} / L_\odot)$ & $12.55_{-0.5}^{+0.3}$ &  $12.25_{-0.4}^{+0.4}$ & $12.65_{-0.4}^{+0.3}$ & $12.55_{-0.5}^{+0.3}$& $12.45_{-0.5}^{+0.4}$\\[0.2cm]

$ T_\mathrm{dust} / K $ & $43_{-10}^{+10}$ &  $38_{-10}^{+10}$ & $43_{-10}^{+10}$ & $38_{-5}^{+15}$ & $43_{-10}^{+10}$\\[0.2cm]

$\log\,(\,M_\mathrm{dust} / M_\odot)$ & $8.75_{-0.4}^{+0.3}$ &  $8.75_{-0.5}^{+0.3}$ & $8.65_{-0.3}^{+0.4}$ & $8.75_{-0.4}^{+0.3}$ & $8.65_{-0.4}^{+0.3}$\\
\enddata
\tablecomments{Average properties of the ALESS galaxies inferred from stacking the likelihood distributions of the physical parameters of individual galaxies. The parameter values listed are the median of the stacked likelihood distribution of each the whole sample of 99 galaxies (first column) and each sub-sample (following columns). The range indicated with each median corresponds to the 16th -- 84th percentile of the likelihood distribution.\vspace{0.25cm}}
\end{deluxetable*}

In order to analyze the overall properties of our sample constrained with \magphys, and at the same time take the uncertainties associated with these constraints into account, in Fig.~\ref{properties_zcut} we stack the all 99 individual likelihood distributions of stellar mass, star formation rate, specific star formation rate, mass-weighted age, $V$-band dust attenuation, $H$-band mass-to-light ratio, total dust luminosity, average dust temperature and total dust mass. We note that in our framework, if the parameters of the galaxies were poorly constrained, these stacked (posterior) likelihood distributions would resemble the prior distributions (Fig.~\ref{priors}) and would change depending on the adopted priors. We tested this by fitting all SEDs with a model library with significantly different priors for the parameters, specifically including a larger fraction of models with older ages, lower specific star formation rates, higher mass-to-light ratios, and a flat redshift distribution. We find that even when using these different parameter priors the resulting stacked posterior distributions and the average properties listed do not change significantly, which indicates that our results are robust, i.e. not biased by the priors.

In Table~\ref{parameters_table}, we list the medians and the 16th--84th percentile ranges of these distributions for each parameter. These ranges reflect the range of values for each parameter found in the sample, taking into account the uncertainties in each individual constraint. 
Overall, the population of ALESS SMGs seems to have fairly uniform properties, with star formation rates, dust luminosities, mass-weighted ages, mass-to-light ratios and dust masses within $\lesssim 1$~dex of the sample median. The range of dust attenuation and temperature values is relatively large (with 16th--84th percentile ranges $0.9 \le A_V \le 3.1$ and $ 33 \le T_\mathrm{dust}/\mathrm{K} \le 53$), but we note that the individual likelihood distributions tend to be broader for these parameters.
To understand if the overall properties vary between SMGs at lower and higher redshifts, we also plot, in Fig.~\ref{properties_zcut}, the contribution to the stacked likelihood distributions from sources with $z_\mathrm{phot} < 2.7$ and from sources with $z_\mathrm{phot} \ge 2.7$ (the median redshift of the sample; \S\ref{redshifts}). The average properties in these two sub-samples are also listed in Table~\ref{parameters_table}.
This shows that the spread in properties shown in Fig.~\ref{properties_zcut} is also due to some extent to the fact that there is a tendency for SMGs at higher redshifts to have higher star formation rates, dust luminosities, dust temperatures and dust attenuation values.
Fig.~\ref{properties_zcut} also shows that our SMGs have a wider range of stellar masses, specific star formation rates and mass-weighted ages in the low-redshift bin ($z<2.7$), with some indication that the distribution of these parameters may even be bimodal. This may be because, at lower redshifts, we may have relatively more massive galaxies with older stellar populations that are also bright in the sub-millimeter. At higher redshifts, we may not be able to detect massive, relatively more `quiescent' galaxies because they have not had time to build up to high stellar masses and older ages. Additionally, since the SEDs of the highest-redshift galaxies are typically less well sampled in the optical than those of the lowest-redshift galaxies, it is possible that in the higher redshift bin any existing structure in the distributions of these parameters may be diluted by the large uncertainties.

\subsection{Stellar masses \& mass-to-light ratios}

\begin{figure*}[t]
\begin{minipage}{0.5\linewidth}
\centering
\includegraphics[width=0.95\textwidth]{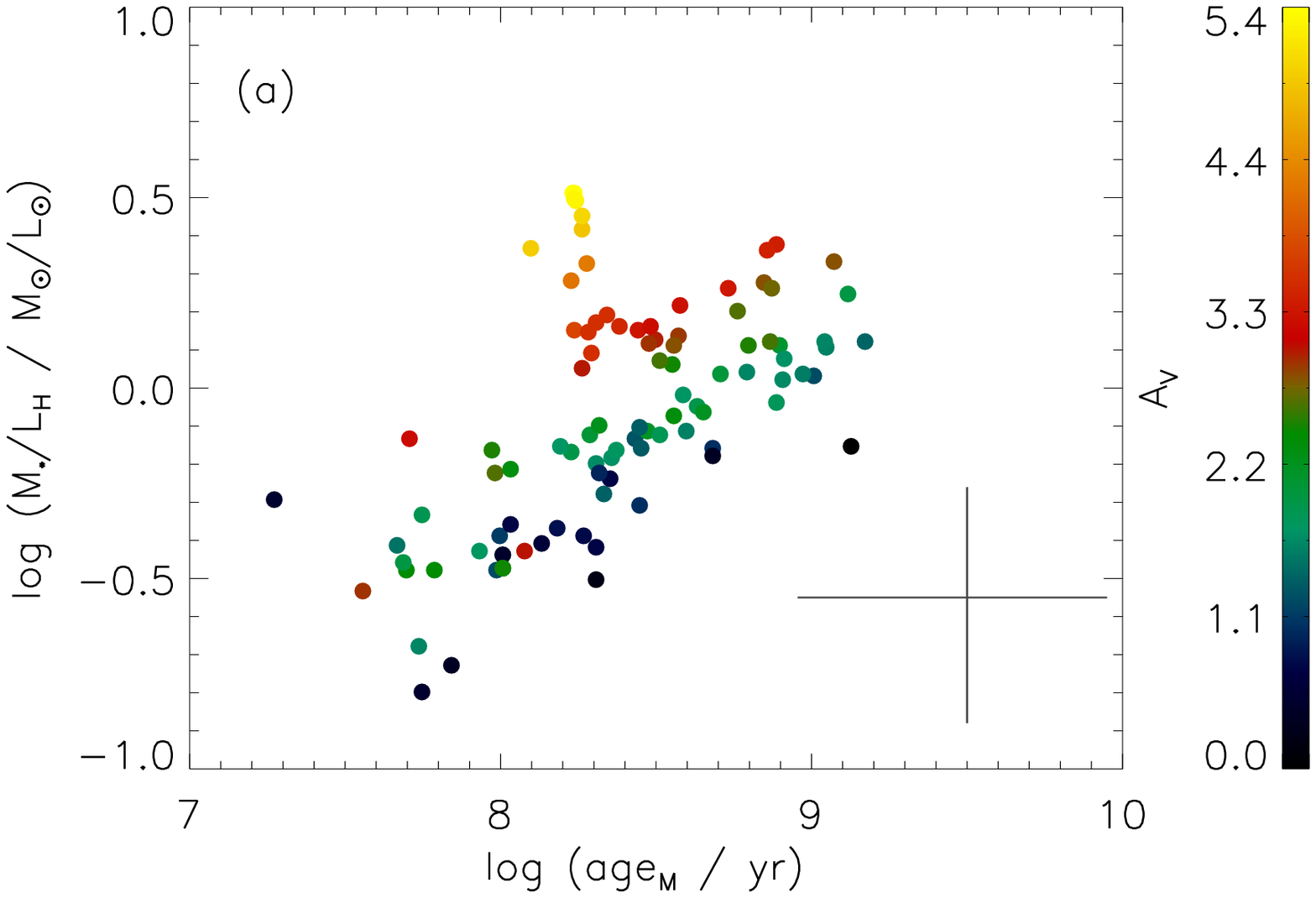}
\end{minipage}
\begin{minipage}{0.5\linewidth}
\centering
\includegraphics[width=0.95\textwidth]{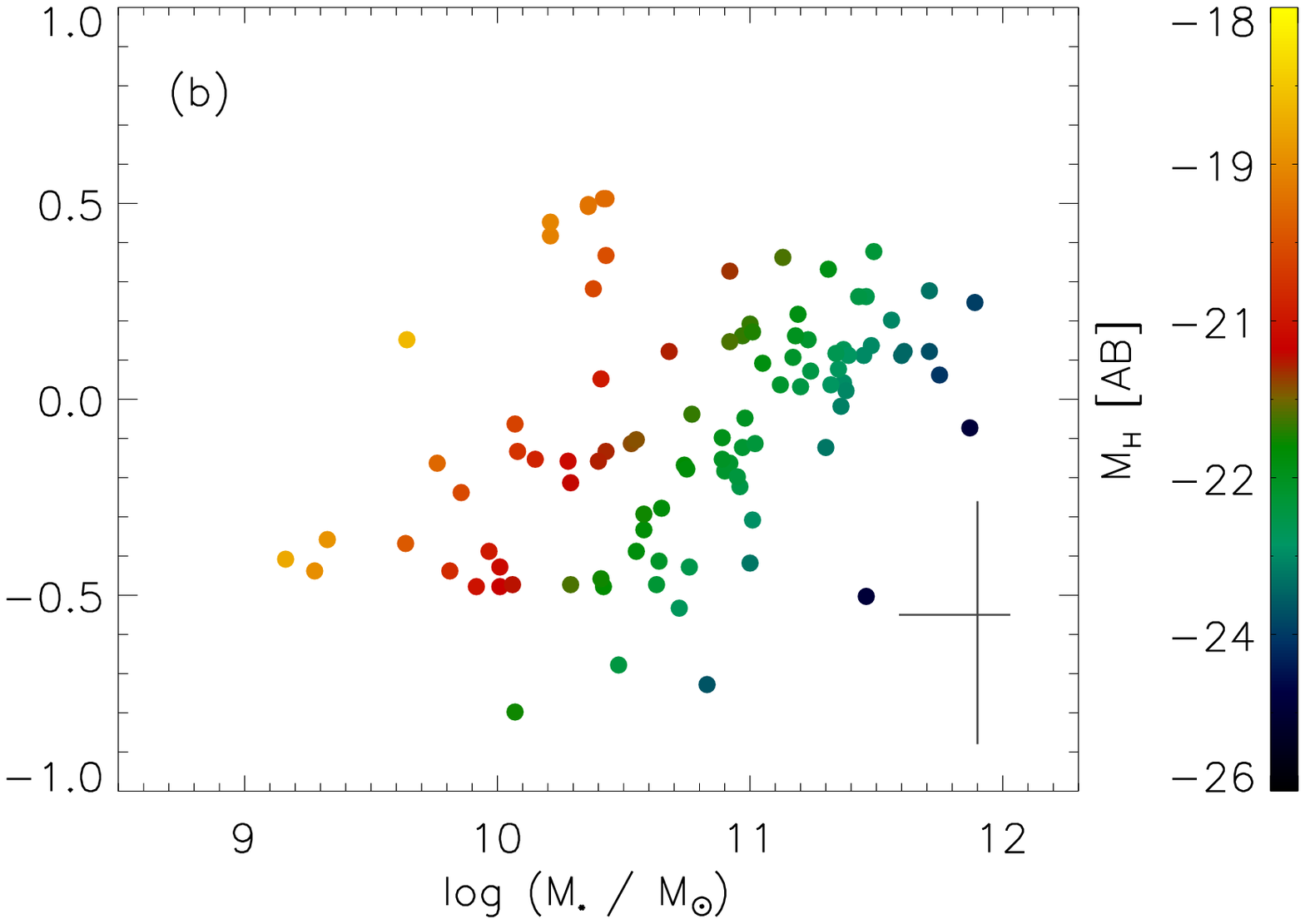}
\end{minipage}
\caption{Median-likelihood estimates of the $H$-band mass-to-light ratio of the 99 ALESS sources plotted against the median-likelihood estimates of (a) the mass-weighted age, (b) the stellar mass. On the bottom right-hand corners, we plot the median error bars (defined as the 16th--84th percentile range of the likelihood distributions). Each galaxy is color-coded according to its derived $V$-band attenuation, $A_V$ (a), and its $H$-band absolute magnitude (b), obtained by dividing the stellar mass of each galaxy with its inferred mass-to-light ratio. We note that the typical error bars on our stellar masses are smaller than for the mass-to-light ratio because we do not use $M_\ast/L_H$ directly to constrain $M_\ast$. In \magphys, each model SED is normalized to 1~\msun, and the stellar mass of a galaxy is constrained by finding the SED normalization that best fits the data using all available bands (as described in \citealt{daCunha2008}).}
\label{mass_light}
\end{figure*}

To understand how well we can constrain the stellar mass of individual SMGs, we analyze their individual stellar mass likelihood distributions\footnote{In this paper `stellar mass' refers to the current mass of stars in the galaxy, i.e. excluding the mass returned to the interstellar medium.}. The stellar mass confidence range of each individual galaxy, given by the 16th -- 84th percentile of the likelihood distribution (marginalized over all other parameters, including redshift), has a median of 0.5~dex. This means that given the uncertainties caused by degeneracies between parameters, unknown star formation histories, sampling of the SEDs, and errors in the photometric data, we constrain the stellar masses of individual objects typically within a factor of 3 (see also \citealt{Hainline2011}). The error bar on each individual stellar mass estimate depends quite strongly on the number of detections in the optical/near-infrared range, which is expected, since the more detections are available, the easier it is to break degeneracies and constrain the parameters. For the optically-faint galaxies, detected in fewer than four optical/near-IR bands, the median confidence range is significantly larger at 0.7~dex (or a factor of 5), while for the galaxies with the most ideal sampling of the optical SED (with more than 10 bands), the median confidence range is 0.3~dex wide (or a factor of 2).

The median stellar mass of the ALESS sources is $8.9\pm0.1\times10^{10}~M_\odot$ (assuming a \citealt{Chabrier2003} IMF), slightly higher than the value found by \cite{Simpson2013} for the same sample ($8\pm1\times10^{10}~M_\odot$ for a Salpeter IMF, which corresponds to about $5\pm1\times10^{10}~M_\odot$ for a Chabrier IMF). The 16th--84th percentile range of the stellar mass stacked likelihood distribution is quite broad, from $\sim 1.4\times 10^{10}$~\msun\ to $\sim 5.6 \times 10^{11}$~\msun, which shows that statistically, there is a wide range of stellar mass values that are consistent with the available data for these galaxies.

The stellar mass confidence ranges become on average $\sim 0.2$~dex broader if only the $U$-band to IRAC photometry is used to constrain the fit (as shown in the likelihood distributions plotted in Fig.~\ref{fit_aless1}). This difference in confidence ranges is particularly pronounced for the optically-faint SMGs, where the stellar emission-only fits yield a median confidence range of 1~dex. This shows that adding the infrared emission constraints in our SED fits can help better constrain even the stellar masses of these galaxies. This is mainly because the infrared information typically reduces the uncertainty in redshift, and it also reduces the uncertainty in dust attenuation $A_V$ (via the constraint on total dust luminosity and energy balance), which affects directly the mass-to-light ratios as shown in Fig.~\ref{mass_light}(a).

We analyze how the stellar mass and mass-to-light ratio estimates from our full SED fits depend on other physical parameters in Fig.~\ref{mass_light}. In Fig.~\ref{mass_light}(a), we plot the $H$-band mass-to-light ratio against the mass-weighted age, which is primarily determined by the star formation history. The mass-to-light ratio correlates with $\mathrm{age}_M$ (and star formation history): the older the ages, the more stellar mass can be `hidden' in old, low-mass stars that contribute little to the light. This has been discussed also by e.g.~\cite{Hainline2011,Simpson2013,Michalowski2014}. The large range of mass-to-light ratios with different star formation histories illustrates why we must marginalize over a wide range of SFHs in our SED fitting to get reliable stellar mass estimates. We find that the typical mass-weighted ages of our sample are less than 1~Gyr (the median for the sample is $\simeq 230\pm140$~Myr), which indicates that the stellar mass is not dominated by a significantly old stellar population (with ages $> 1$~Gyr) as argued by \cite{Michalowski2012}. We note that, even though the mass-weighted age is a very hard parameter to constrain, this result does not seem to be driven by our modelling assumptions, since our model library includes models with $\mathrm{age}_M > 10^{9}$~yr, as can be seen in the prior histogram plotted in Fig.~\ref{priors}(a). The crucial difference between our analysis and that of \cite{Michalowski2012} is that we include a wide range of star formation histories, ages, metallicities, dust attenuations and redshifts, and when we fit the whole ultraviolet-to-radio SED we require self-consistency between the stellar and dust emission (in terms of energy balance, as detailed in Section~\ref{combined}). When doing so, it becomes clear that these galaxies must have a significant amount of ongoing star formation to power the large infrared luminosities, and even when allowing for models to have a significant fraction old stars (our star formation history prior also includes models with early bursts/peaks of star formation, which have older mass-weighted ages), that is not what is preferred by the fitting in a statistical sense, i.e. the posterior likelihood distributions of the age are not peaking at the oldest ages allowed by the prior. If indeed the data preferred older ages, we would see that the posterior likelihood distributions would peak at $\log(\mathrm{age_M})>9$ and even towards the edge of allowed ages.
The relation between mass-to-light ratio and age is broadened by dust attenuation even in the $H$ band: Fig.~\ref{mass_light}(a) shows that, at fixed age, the higher $A_V$, the higher the $H$-band stellar mass-to-light ratio. This shows how important it is to properly model the effects of dust in the stellar SEDs via our energy balance technique to better constrain $A_V$ and hence the mass-to-light ratios and stellar masses.

In Fig.~\ref{mass_light}(b), we plot the $H$-band mass-to-light ratio against our stellar mass estimates, where each galaxy is color-coded by its rest-frame $H$-band absolute magnitude (computed by dividing the stellar mass of each galaxy by its inferred mass-to-light ratio). The median $M_H$ inferred from our SED fits is $M_H=-22.9\pm1.2$. At fixed $M_H$, the stellar mass correlates tightly with the mass-to-light ratio, and we find that the highest mass galaxies have higher $M_\ast/L_H$, meaning they have possibly a larger number of older stars and/or higher dust attenuation.

\subsection{Star formation rates}

\begin{figure}
\begin{center}
\vspace{0.2cm}
\includegraphics[width=0.475\textwidth]{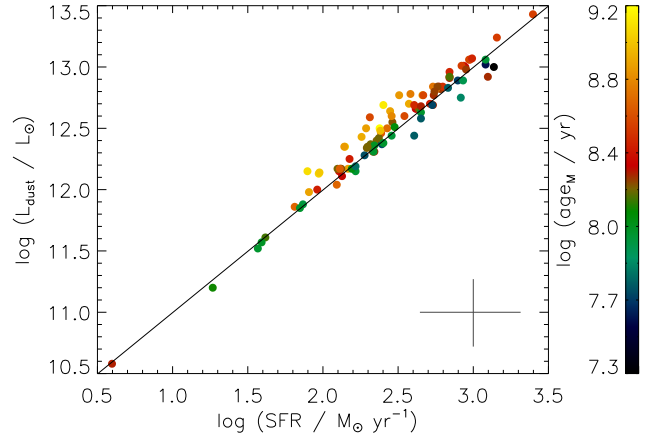}
\caption{Median-likelihood estimates of the star formation rate against the total dust infrared luminosity of the ALESS sources. The black line shows the relation of \cite{Kennicutt1998} (scaled to a \cite{Chabrier2003} IMF) which essentially assumes an optically-thick starburst.}
\label{sfr_ldust}
\end{center}
\end{figure}

We find a median star formation rate for the ALESS sources of $280\pm70~\msun~\mathrm{yr}^{-1}$, which is consistent with the value inferred by \cite{Swinbank2013} using the total infrared luminosity ($300\pm30~\msun~\mathrm{yr}^{-1}$). The star formation rates are consistent because, as shown in Fig.~\ref{sfr_ldust}, the total dust luminosity traces the star formation rate extremely well (within $\sim0.1$~dex), in the same way as predicted using the \cite{Kennicutt1998} standard conversion between infrared luminosity and SFR (plotted as a black line). This is expected for  very dusty, actively star-forming sources for which the \cite{Kennicutt1998} conversion is best calibrated (as also discussed by e.g.~\citealt{Kennicutt2009,daCunha2010b,Rowlands2014}). Fig.~\ref{sfr_ldust} also shows that galaxies with older mass-weighted ages typically have slightly higher dust luminosity per unit star formation rate (up to $\sim0.15$~dex) than predicted by the \cite{Kennicutt1998} conversion. This is a consequence of additional dust heating by relatively old stellar populations (that are more important in galaxies with higher $\mathrm{age}_M$), which increases the dust luminosity at fixed star formation rate.

The very tight correlation between the infrared luminosity and the star formation rate that we find is not directly imposed in \magphys, since the model allows for significant heating of dust by old stellar populations \citep{daCunha2008}, which could imply larger dust luminosities per unit star formation rate than those predicted by the \cite{Kennicutt1998} conversion or, in other words, the fit would need lower SFRs to reproduce the observed dust luminosity. While models with non-negligible heating of dust by older stellar populations are allowed in the libraries we use to fit the data, the results indicate that the observed SEDs (including energy balance) constrain our sources to be very dusty, actively star-forming galaxies where the dust emission is tracing the SFR.

\subsection{Dust properties}
\label{section_dust}

\begin{figure}
\begin{center}
\vspace{0.2cm}
\includegraphics[width=0.475\textwidth]{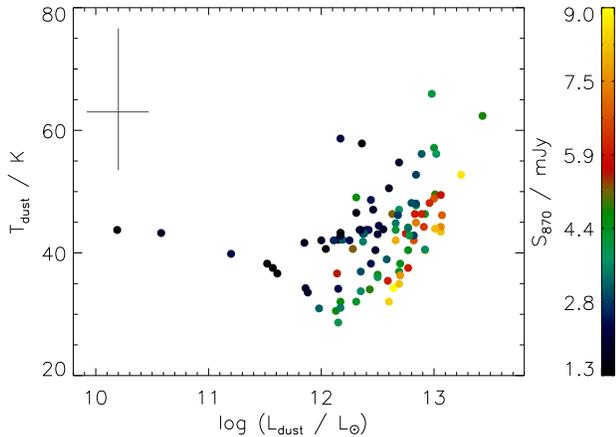}
\caption{Luminosity-averaged dust temperature of our SMGs as a function of their total dust luminosity. Each SMG is color-coded according to its measured ALMA 870-\mic\ flux. In the top left-hand corner we plot the median error bar on these properties, defined as the 16th -- 84th percentile range of the likelihood distributions.}
\label{ldust_tdust}
\end{center}
\end{figure}

On average, the typical ALESS source has a dust luminosity of $(3.5\pm0.8)\times10^{12}~\lsun$, a dust mass of $(5.6\pm1.0)\times10^{8}~\msun$, and a luminosity-averaged dust temperature of $43\pm2$~K.

As previous other SMG studies (e.g.~\citealt{Chapman2005,Kovacs2006,Wardlow2010,Magnelli2012,Symeonidis2013,Swinbank2013, Casey2014}), we find a correlation between the total dust luminosity and the average temperature of our sources (plotted in Fig.~\ref{ldust_tdust}). The measured dust temperatures in \cite{Chapman2005} and \cite{Swinbank2013} are typically lower than our luminosity-averaged temperatures, but this could be simply because of the different temperature definition and/or the way the infrared SEDs are fitted (as discussed in e.g.~\citealt{Casey2012,Hayward2012,Magnelli2012}).
Our luminosity-temperature correlation is broader than that observed for the SMGs of \cite{Chapman2005} and \cite{Kovacs2006} because our sample includes lower sub-millimeter flux sources thanks to the high sensitivity of ALMA. At fixed infrared luminosity, the deeper we go in sub-millimeter flux $S_{870\mic}$, the higher dust temperatures (and lower dust mass) we probe. This means that SMGs only have very cold dust temperatures (compared to local galaxies of similar infrared luminosity) if we consider the sources detected at the highest sub-millimeter fluxes (see also \citealt{Magnelli2012}).

We find a weak correlation between the total dust luminosity (and the average dust temperature) and the average $V$-band dust attenuation $A_V$. If we divide our sample into sources with dust attenuation lower than the sample median (i.e. $A_V < 1.9$) and sources with dust attenuation higher than the sample median ($A_V > 1.9$), we find that galaxies with the lowest $A_V$ have a median dust temperature $T_\mathrm{dust}=38\pm2$~K and a median  dust luminosity $L_\mathrm{dust}=(2.2\pm0.8)\times10^{12}~\lsun$, while the highest $A_V$ galaxies have a median dust temperature  $T_\mathrm{dust}=42\pm2$~K and dust luminosity $L_\mathrm{dust}=(4.6\pm0.8)\times10^{12}~\lsun$. While this is a very tentative result given the large error bars on our $A_V$ and $T_\mathrm{dust}$ estimates, it may indicate that, in more dust-obscured galaxies, dust is more effectively heated, presumably because of a more compact distribution of dust, i.e. the higher $A_V$, the more concentrated dust grains are around the (young) stars, and so they see a stronger radiation field (we return to this discussion in \S\ref{section_average_seds}). A better sampling of the SEDs (possibly combined with known redshifts) is needed to reduce the statistical error bars on these parameters in order to investigate this correlation in more detail.

\section{Discussion}
\label{discussion}

\subsection{The average spectral energy distribution of ALESS SMGs}
\label{section_average_seds}

\begin{figure*}
\centering
\includegraphics[width=0.75\textwidth]{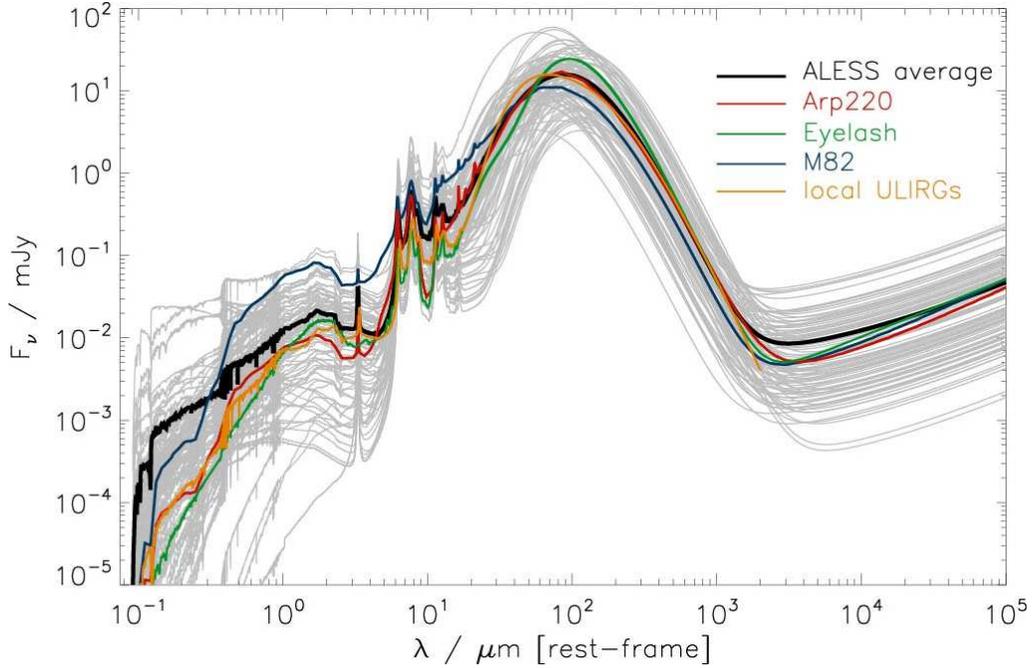}
\vspace{0.2cm}
\caption{Best-fit model spectral energy distributions of the 99 ALESS sources obtained from our fits. In grey, we plot the SED of each individual source (we use our photometric redshift estimate to plot each SED in the rest-frame). We also plot the average SED obtained by averaging the flux $F_\nu$ of each of the 99 best-fit model SEDs at each (rest-frame) wavelength. For comparison, we plot the template SEDs of the `prototypical' local ULIRG Arp220, the local starburst galaxy M82 \citep{Silva1998}, the average SED of 16 local ULIRGs from \cite{daCunha2010b}, and the $z=2.3$ sub-millimeter galaxy SMMJ2135-0102 (`Cosmic Eyelash'; \citealt{Swinbank2010}). These four templates are normalized such that their total infrared luminosity is the same as the median total infrared luminosity of the whole sample i.e. $3.5\times10^{12}~\lsun$, at the median redshift $z=2.7$.}
\centering
\label{average_seds}
\end{figure*}

\begin{figure}
\centering
\includegraphics[width=0.5\textwidth]{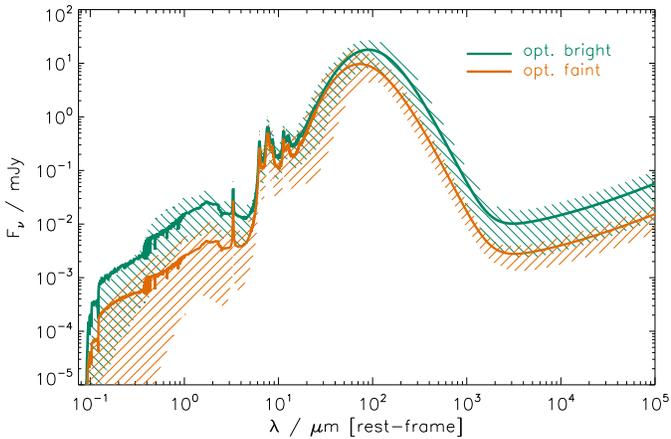}
\caption{Comparison between the average SED of the 77 optically-bright sources and the 22 optically-faint sources. The shaded regions show the 16th to 84th percentile range for each subsample.}
\centering
\label{average_seds_bright_faint}
\end{figure}

\begin{figure*}
\centering
\includegraphics[width=\textwidth]{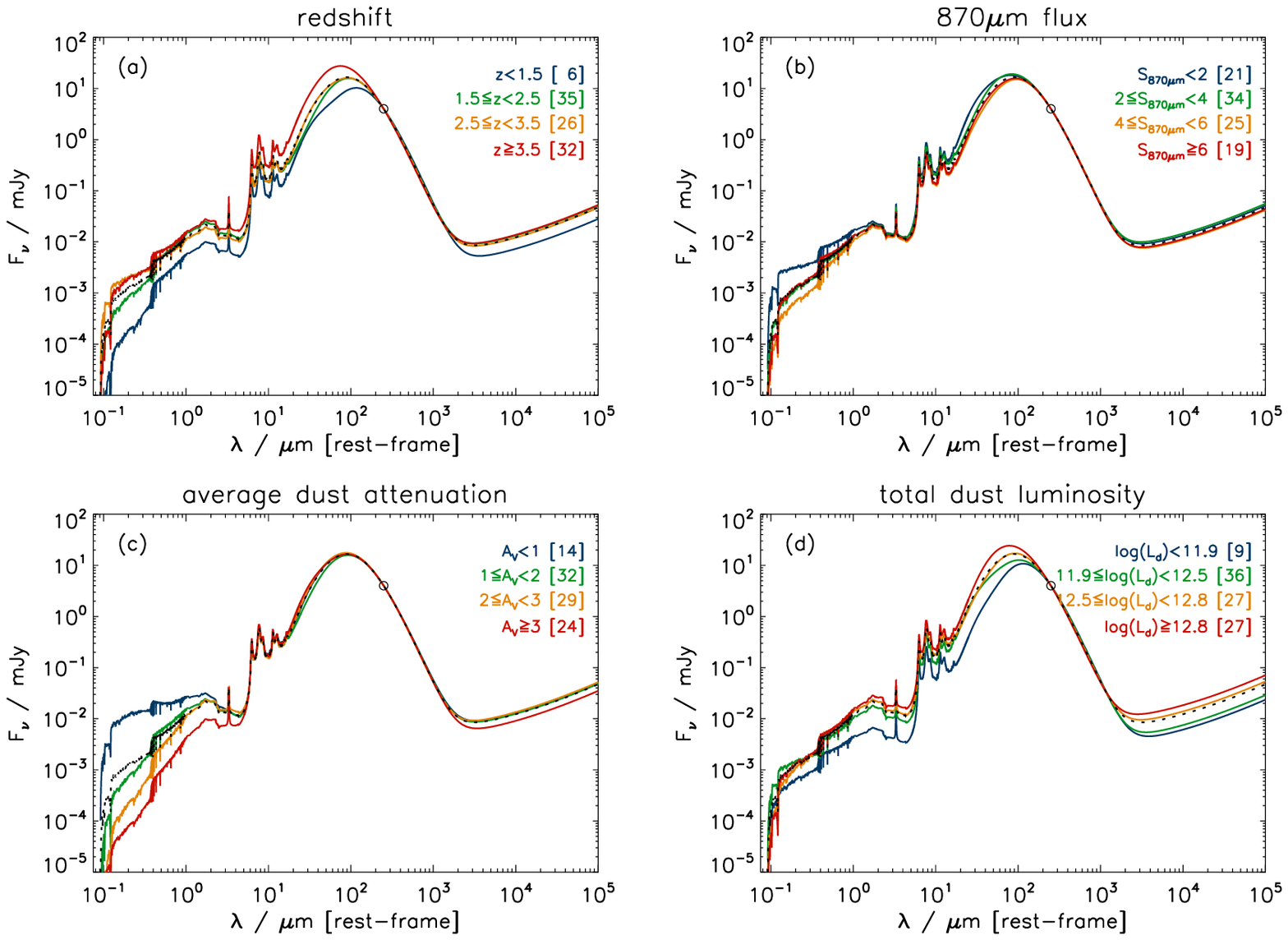}
\caption{Variation of the intrinsic SED shapes of the 99 ALESS SMGs with different properties: (a) redshift; (b) observed 870-\mic\ flux measured with ALMA (in mJy); (c) average $V$-band dust attenuation; (d) total dust infrared luminosity (in \lsun). In each panel, we plot the average SED in four different bins of the parameter being considered, indicated in the top right-hand corner (the number in brackets indicates the number of sources falling in each bin). For reference, we also plot the average SED of all the sources (shown in Fig.~\ref{average_seds}(a)) as a black dotted line. All the SEDs are normalized to the average ALMA 870\mic\ flux of the sample, 4 mJy, at rest-frame 250\mic, which corresponds to observed wavelength 870\mic\ for a source at $z=2.5$ (indicated by the black circle).}
\label{average_seds_bins}
\end{figure*}

In this section we analyze the best-fit \magphys\ ultraviolet-to-radio SEDs of our ALESS SMGs.
In Fig.~\ref{average_seds}, we plot the best-fit model SEDs of all 99 main ALESS SMGs in the rest-frame (calculated using our photometric redshift for each SMG), in observed flux units, highlighting the range of intrinsic SEDs of these sources. We obtain the `average ALESS SED' (plotted in black in Fig.~\ref{average_seds}) by computing the simple average flux $F_\nu$ of all the best-fit SEDs at each rest-frame wavelength (weighting the average by infrared luminosity or sub-millimeter flux does not change the result). We also plot, for comparison, the SEDs of two prototypical local starburst galaxies, Arp220 (a ULIRG) and M82 (a dwarf starburst), the average SED of 16 local ULIRGs from \cite{daCunha2010b}, and the $z=2.3$ sub-millimeter galaxy SMMJ2135-0102 (the `Cosmic Eyelash'; \citealt{Swinbank2010}), all normalized such that their total infrared luminosity is the same as the median total infrared luminosity of the whole sample i.e. $3.5\times10^{12}~\lsun$, at the median redshift $z=2.7$. The spectral energy distribution of the average ALESS SMG in the infrared ($\lambda \gtrsim 5$\mic) follows closely that of Arp220 (with the exception of the 9.8-\mic\ silicate absorption feature that is included in the Arp220 template but not in the \magphys\ models). In the rest-frame ultraviolet to near-infrared range, both the Arp220 template and the average local ULIRG template are fainter and redder than the average ALESS SMG, which could be due to a different stellar content and/or dust attenuation. This is consistent with high-redshift SMGs being more massive and/or having a more extended dust distribution than local ULIRGs. While Arp220 may provide a suitable template for the (rest-frame infrared) dust emission of high-redshift SMGs, caution must be taken when using this template to interpret or extrapolate their (rest-frame UV-to-near-IR) stellar emission. Fig.~\ref{average_seds} also shows that the average emission by ALESS SMGs is very different from the starburst M82, which shows hotter dust emission, significantly higher optical/near-infrared flux, and a redder stellar continuum. It is clear from this comparison that the full SED of high-redshift SMGs is not simply a scaled-up version of a local starburst. Interestingly, \cite{Delmestre2009} find that the mid-infrared spectral properties of SMGs are similar to those of local starbursts, but due to the poor sampling of the rest-frame mid-infrared emission, we cannot verify whether this is also the case for the ALESS SMGs.

In Fig.~\ref{average_seds}, we compare the average SEDs of our 77 optically-bright sources and our 22 optically-faint sources. Not surprisingly, the ultraviolet-to-near-infrared emission of the optically-faint sample is significantly fainter (about an order of magnitude) than for the optically-bright sample, and the slope of the stellar continuum is also redder. This is consistent with the significantly higher inferred $A_V$ of these SMGs as discussed in \S\ref{section_faint}. Interestingly, the average infrared SED of these SMGs peaks at shorter wavelength than the optically-bright average SED, which could indicate that these sources have on average higher dust temperatures.
This tentative evidence for warmer dust in the optically-faint sources, combined with higher inferred dust attenuation $A_V$ may indicate that these galaxies are more compact than their optically-bright counterparts, which would increase the dust column (hence higher $A_V$) and provide more effective dust heating by the stellar radiation field (e.g.~\citealt{Elbaz2011}).

To understand the variation of intrinsic SEDs of the 99 ALESS SMGs, we plot, in Fig.~\ref{average_seds_bins}, the average SEDs in different bins of (a) redshift, (b) observed ALMA 870-\mic\ flux, (c) average $V$-band dust attenuation $A_V$, and (d) and total dust luminosity. We note that the scatter within each bin is large and in some cases larger than the difference between the bins, but nevertheless this is a useful way of understanding average variations of SED shapes as a function of different properties. A larger sample with more complete sampling of the SEDs for all sources is needed to perform a more quantitative statistical analysis.

Fig.~\ref{average_seds_bins}(a) shows that the average intrinsic SEDs of the ALESS SMGs are different depending on their redshift: as redshift increases, we select intrinsically brighter sources with that peak at lower wavelengths in the (rest-frame) infrared i.e. have warmer dust. Fig.~\ref{average_seds_bins}(b) shows that the average intrinsic SEDs of SMGs of different observed 870-\mic\ flux do not vary significantly: this does not mean that the SEDs do not vary within each flux bin, but rather that there is no systematic variation with sub-millimeter flux, i.e. we are not selecting a particular type of SED in different flux bins. If all we know about an SMG is its 870-\mic\ flux and redshift, then the average ALESS SED is a fair approximation for its intrinsic SED regardless of its sub-millimeter flux.
Fig.~\ref{average_seds_bins}(c), illustrates how the average dust attenuation is a crucial parameter driving the variability of optical SED shapes in the ALESS sample; as expected, as $A_V$ increases, the stellar continuum of the galaxies becomes systematically fainter and redder. We find that the infrared emission in different $A_V$ bins remains almost constant, with only a small increase of the total infrared luminosity and dust temperature towards higher $A_V$. The fact that the infrared SEDs vary little with $A_V$ is a possible indication that the main difference between galaxies in these bins is viewing angle (i.e.~galaxies with higher $A_V$ being observed more edge-on), since viewing angle does not affect the shape and normalization of the dust emission because it is isotropic (for an illustration of the effect of viewing angle on SEDs computed using a radiative transfer code, see figure 8 of \citealt{Jonsson2010}). The small observed variation in the shape and normalization of the dust emission in different bins (in the sense that higher $A_V$ sources are on average slightly hotter and more infrared-luminous) could be due to variations on the compactness of the sources, which as discussed above can be driving the difference between the SEDs of optically-bright and optically-faint sources. In this scenario, in more compact galaxies, the dust column is higher (hence higher $A_V$), and the dust gets more effectively heated because it is closer to the heating source (the stars) and hence feels a stronger radiation field.
In Fig.~\ref{average_seds_bins}(d), we plot the average SEDs in bins of total dust luminosity. The peak of the dust emission shifts to lower wavelengths as the dust luminosity increases, which reflects the luminosity-temperature relation discussed in \S\ref{section_dust}. We also note that the average stellar continuum gets redder as the dust luminosity increases. This is mainly caused by the fact that the highest dust luminosity sources have higher stellar masses (and thus stronger near-infrared stellar bumps) and also, to a lesser extent, because they have slightly higher average $A_V$.

\subsection{The nature of SMGs: comparison with the `star-forming main sequence'}

\begin{figure*}
\centering
\includegraphics[width=\textwidth]{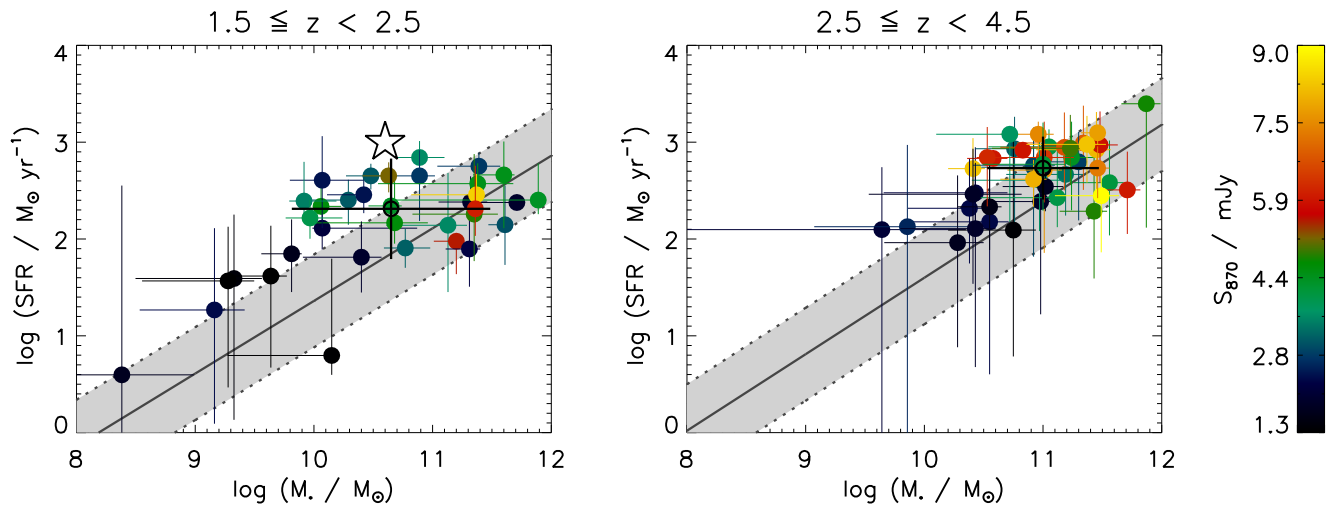}
\vspace{-5.5cm}
\caption{Comparison of the stellar masses and star formation rates of our ALESS SMGs with the `star-forming main sequence' in two different redshift bins: $1.5 \le z < 2.5$ (left) and $2.5 \le z < 4.5$ (right). Each galaxy is colour-coded by its 870-\mic\ flux density measured from the ALMA observations \citep{Hodge2013}. The dark grey solid lines show the position of the star-forming main sequence at each redshift $z=2$ (left) and $z=3.5$ (right) as given by \cite{Speagle2014}, based on a compilation of different studies in the literature spanning a wide range in stellar masses and redshifts. The star formation rate of main sequence galaxies is given by: $\log(\mathrm{SFR}_\mathrm{MS}(z,M_\ast)/M_\sun~\mathrm{yr}^{-1})=(0.84-0.026\,t(z))\log(M_\ast/M_\sun)+0.11t(z)-6.51$, where $t(z)$ is the age of the Universe at redshift $z$ in Gyr. The grey dotted lines indicate a factor of 3 above and below this main sequence.
For reference, the black star shows the `average SMG' of \cite{Daddi2007} (with $\log(M_\ast/M_\odot)=10.6$ and $\log(\mathrm{SFR}/M_\odot~\mathrm{yr}^{-1})=3$). The black open circle in each panel shows the median stellar mass and star formation rate of the ALESS sources in that redshift range (with the error bar indicating the standard deviation).}
\vspace{1.5cm}
\label{ms}
\end{figure*}

In this section we discuss, based on our results for the physical properties of the ALESS SMGs, how they compare with the `normal' galaxy population at their redshifts. The place occupied by SMGs in the star formation rate vs. stellar mass plane is a subject of great ongoing discussion in the literature (e.g.~\citealt{Hainline2011,Magnelli2012,Michalowski2012}).
Some studies argue that SMGs are extreme starbursts forming stars at much higher rates than other galaxies of the same stellar mass at the same redshift, implying that these intense starbursts have to be powered by extreme events such as major merger (\citealt{Daddi2007,Engel2010,Hainline2011,Magnelli2012}), while some cosmological simulations establish SMGs simply as the high-mass end of the normal galaxy population, implying that whatever `secular' processes govern the star formation histories of fainter galaxies are also at play in the sub-millimeter bright population \citep{Dave2010}.

Our careful analysis of the star formation rate and stellar mass of the ALESS sample (a complete and unbiased SMG sample) gives us a unique opportunity to address this question. In order to compare the ALESS SMGs with the general galaxy population at their redshifts, in Fig.~\ref{ms} we plot the star formation rates against stellar masses of our sources in two redshift bins ($1.5 \le z < 2.5$ and $2.5 \le z < 4.5$), and compare them with the `star-forming main sequence' (i.e. the observed correlation between the stellar mass and the star formation rate of the general galaxy population) at those redshifts. The exact slope, normalisation and redshift evolution of the main sequence still have large uncertainties and vary between studies; here we chose the main sequence definition of \cite{Speagle2014}, which is based on a compilation of different studies in the literature spanning a wide range of stellar masses and redshifts. We plot their main sequence relation at $z=2.0$ and $z=3.5$ as a grey solid line in the left- and right-hand panels of Fig.~\ref{ms}, respectively.

Fig.~\ref{ms} shows that, contrary to less-active galaxies in mass-selected samples, there is no strong correlation between the stellar masses and star formation rates of the ALESS sources at fixed redshift: this is because the sub-millimeter selection essentially selects in star formation rate. We find that, at $z\simeq2$, our SMGs have on average $M_\ast\simeq4.5\times10^{10}~M_\sun$ and SFR$\simeq205~M_\sun~\mathrm{yr}^{-1}$, which implies that their specific star formation rate is about 0.7~dex lower than that of the `average $z=2$ SMG' of \cite{Daddi2007} (plotted for comparison). With the relatively more modest specific SFRs of our SMGs, we find that $17\pm2$ (i.e.$\sim49$\%) of them lie significantly above the main sequence at that redshift (i.e. they have star formation rates over a factor of 3 higher than `main sequence' galaxies of the same stellar mass), while $18\pm2$ ($\sim51$\%) are consistent with being in the high-mass end of the main sequence.
At higher redshift ($2.5<z<4.5$), the star formation rates and stellar masses of the SMGs are typically higher than at $z\simeq2$, with $M_\ast\simeq10^{11}~M_\sun$ and SFR$\simeq540~M_\sun~\mathrm{yr}^{-1}$ on average. However, since the specific star formation rate of main sequence galaxies predicted by \cite{Speagle2014} continues increasing with redshift, this means that a smaller fraction of our SMGs lie significantly above the main sequence at those redshifts. We find that only $11\pm2$ SMGs (i.e.$\sim27$\%) have SFRs at least 3 times higher than main sequence galaxies of the same stellar mass at the same redshift, with most SMGs (73\%) being consistent with the main sequence. Our results depend on how well the evolution of the star-forming main sequence with redshift is constrained. If the specific star formation of main sequence galaxies plateaus at $z>2$ as suggested by some studies (e.g.~\citealt{Weinmann2011,Gonzalez2014}), then the fraction of high-redshift SMGs that are significant outliers would increase. The wide stellar masses and positions with respect to the star-forming main sequence of the ALESS SMGs may be an indication that these galaxies are not a uniform population (as suggested by e.g.~\citealt{Hayward2012}).

\section{Summary \& Conclusions}
\label{conclusion}

In this paper we have presented an update of the \magphys\ code that allows us to fit simultaneously the ultraviolet-to-radio SEDs of an unbiased sample of SMGs from the ALESS survey. This new version of the code allows us to constrain photometric redshifts simultaneously with all other physical properties, and explores a realistic parameter space for high-redshift star-forming galaxies, including complex star formation histories, and a wide range of dust attenuation and emission parameters to include different possible dust contents and distributions. Our main results are summarized as follows.

\begin{itemize}

\item Our redshifts are consistent with the classic method of deriving photometric redshifts by fitting only the stellar emission (i.e. no dust emission). Our method improves the redshift constraints in the case of optically-faint sources for which optical observations are limited, by including information from infrared and radio wavelengths. An advantage of our method is that the uncertainty in photometric redshift is naturally included in the uncertainty in the intrinsic physical parameters of the galaxies such as the stellar mass, dust attenuation, dust luminosity, dust temperature and star formation rate (and vice-versa).

\item We derive the median properties of the full sample of 99 ALESS SMGs by has a median redshift $z=2.7\pm0.1$, has a median stellar mass of $\simeq8.9\pm0.1\times10^{10}$~\msun, a star formation rate of $\simeq300~\msun~\mathrm{yr}^{-1}$, average $V$-band dust attenuation $A_V \simeq2$, total dust luminosity
$L_\mathrm{dust}=\simeq4\times10^{12}$~\lsun, total dust mass $M_\mathrm{dust}\simeq6\times10^{8}$~\msun, and luminosity-averaged dust temperature $T_\mathrm{dust}\simeq43$~K. These physical properties are very similar to the properties of local ULIRGs \citep{daCunha2010b}.

\item We find that the 22 optically-faint ALESS SMGs are likely to be at higher redshifts than the optically-bright sources, with a median photometric redshift of $z_\mathrm{phot}=3.7$. Our analysis of the likelihood distributions of the physical parameters of these galaxies indicates that they are consistent with having similar overall properties as the bright sources, except for significantly higher dust attenuation. This could indicate that these SMGs are a population of high-redshift SMGs that are either very compact or very edge-on compared to sources of similar sub-millimeter fluxes that are more optically-bright.

\item Using our multi-wavelength SED fits, we provide new SMG templates that should be more appropriate to interpret and/or extrapolate SMG observations than current local galaxy templates.

\item When we compare the star formation rates and stellar masses of the ALESS SMGs with the star formation main sequence at that redshift, we find that, at $z\simeq2$, about half of the SMGs lie significantly above the main sequence (with SFRs 3 times higher than main sequence galaxies of the same stellar mass), while the other half falls in the high-mass end of the main sequence.
At higher redshifts ($z\simeq3.5$), the SMGs tend to have higher SFRs and stellar masses than at $z\simeq2$, but if we include the evolution of the star-forming main sequence with redshift \citep{Speagle2014}, we find that only about a third of the SMGs can be considered extreme, high-SFR outliers. This suggests that the ALESS SMGs are not a uniform population, including galaxies that may be extreme starbursts but also galaxies with SFRs that are similar to those of the general population of galaxies at the same redshift. However, in order to fully understand what this means in terms of the star formation mode of SMGs, we need to more detailed information such as spatially-resolved imaging and spectroscopy (to determine, for example, their dynamics and merging state), and we also need to understand the processes that drive the star-forming main sequence at high redshifts in more detail.

\end{itemize}

Our method is designed to optimally extract as much information as possible on the physical properties of SMGs (or, more generally, any sample of high-redshift star-forming galaxies with a wide range of dust contents and properties) using integrated multi-wavelength observations. We will make the new \magphys\ model library, as well as our average SMG templates, available to the community on the \magphys\ website (www.iap.fr/magphys).


\section*{Acknowledgements}

We thank the referee for their careful reading of the manuscript and their useful and constructive comments. EdC thanks Chris Hayward, Brent Groves and Camilla Pacifici for useful discussions, and acknowledges funding through the ERC grant `Cosmic Dawn'.
IRS acknowledges support from STFC (ST/L00075X/1), the ERC Advanced Investigator grant `DustyGal' 321334, and a Royal Society/Wolfson Merit Award.
JMS and ALRD acknowledge the support of STFC studentships (ST/J501013/1 and ST/F007299/1, respectively).
AMS acknowledges support from an STFC Advanced Fellowship (ST/A005234/1).
RD acknowledges funding from Germany's national research center for aeronautics and space (DLR, project FKZ 50 OR 1104).
AK acknowledges support by the Collaborative Research Council 956, sub-project A1, funded by the Deutsche Forschungsgemeinschaft (DFG).
We acknowledge the hospitality of the Aspen Center for Physics, which is supported by the National Science Foundation under Grant No. PHYS-1066293.


\def\aj{AJ}
\def\araa{ARA\&A}
\def\apj{ApJ}
\def\apjl{ApJ}
\def\apjs{ApJS}
\def\apss{Ap\&SS}
\def\aap{A\&A}
\def\aapr{A\&A~Rev.}
\def\aaps{A\&AS}
\def\mnras{MNRAS}
\def\pasp{PASP}
\def\pasj{PASJ}
\def\qjras{QJRAS}
\def\nat{Nature}

\def\aplett{Astrophys.~Lett.}
\def\aas{AAS}
\let\astap=\aap
\let\apjlett=\apjl
\let\apjsupp=\apjs
\let\applopt=\ao


\appendix

\section{Physical parameter estimates of each ALESS SMG}
\label{appendix_parameters}

\begin{small}
\begin{deluxetable*}{lrrrrrrrrr}
\tablecolumns{5}
\tablewidth{0pt}
\tablecaption{Median-likelihood estimates (and confidence ranges) of several physical parameters for each ALESS SMG \label{parameters_table2}}
\tablehead{
\colhead{ID} &
\colhead{$z_\mathrm{phot}$} &
\colhead{$\log(M_\ast/M_\sun)$} &
\colhead{$\log(\mathrm{SFR}/M_\sun \mathrm{yr}^{-1})$} &
\colhead{$\log(\mathrm{age}_M/\mathrm{yr})$} &
\colhead{$A_V$} &
\colhead{$\log(M_\ast/L_\mathrm{H})$} &
\colhead{$\log(L_\mathrm{dust}/L_\sun)$} &
\colhead{$T_\mathrm{dust}/K$} &
\colhead{$\log(M_\mathrm{dust}/M_\sun)$}
\\
\\
\colhead{(1)} &
\colhead{(2)} &
\colhead{(3)} &
\colhead{(4)} &
\colhead{(5)} &
\colhead{(6)} &
\colhead{(7)} &
\colhead{(8)} &
\colhead{(9)} &
\colhead{(10)}
}
\startdata
ALESS001.1 & $4.78^{+2.65}_{-1.81}$ & $10.97^{+0.15}_{-0.42}$ & $2.78^{+0.49}_{-0.55}$ & $8.29^{+0.47}_{-0.64}$ & $2.0^{+1.4}_{-1.1}$ & $-0.12^{+0.46}_{-0.52}$ & $12.82^{+0.43}_{-0.45}$ & $42^{+20}_{-14}$ & $9.12^{+0.49}_{-0.46}$ \\[0.1cm] 
ALESS001.2 & $5.22^{+2.21}_{-2.55}$ & $10.90^{+0.13}_{-0.39}$ & $2.62^{+0.45}_{-0.73}$ & $8.36^{+0.51}_{-0.58}$ & $1.7^{+1.9}_{-0.9}$ & $-0.18^{+0.59}_{-0.41}$ & $12.66^{+0.39}_{-0.57}$ & $45^{+21}_{-16}$ & $8.69^{+0.61}_{-0.37}$ \\[0.1cm]  
ALESS001.3 & $3.12^{+1.10}_{-1.85}$ & $10.41^{+0.41}_{-0.88}$ & $2.46^{+0.47}_{-0.92}$ & $8.26^{+0.73}_{-0.72}$ & $3.0^{+1.7}_{-1.5}$ &  $0.05^{+0.61}_{-0.52}$ & $12.55^{+0.39}_{-0.88}$ & $44^{+20}_{-17}$ & $8.39^{+0.44}_{-0.24}$\\[0.1cm] 
ALESS002.1 & $3.12^{+0.41}_{-1.70}$ & $10.72^{+0.01}_{-0.62}$ & $3.08^{+0.08}_{-0.90}$ & $7.56^{+0.81}_{-0.14}$ & $2.9^{+0.7}_{-0.8}$ & $-0.53^{+0.52}_{-0.08}$ & $13.02^{+0.14}_{-0.85}$ & $56^{+6}_{-24}$ &  $8.68^{+0.20}_{-0.18}$ \\[0.1cm] 
ALESS002.2 & $3.78^{+1.69}_{-1.06}$ & $11.00^{+0.19}_{-0.51}$ & $2.77^{+0.43}_{-0.48}$ & $8.34^{+0.53}_{-0.68}$ & $3.5^{+1.4}_{-1.5}$ &  $0.19^{+0.49}_{-0.54}$ & $12.84^{+0.36}_{-0.38}$ & $48^{+18}_{-14}$ & $8.64^{+0.35}_{-0.22}$ \\[0.1cm] 
ALESS003.1 & $3.88^{+0.90}_{-0.76}$ & $11.37^{+0.16}_{-0.51}$ & $2.98^{+0.29}_{-0.40}$ & $8.50^{+0.34}_{-0.70}$ & $3.0^{+0.8}_{-0.9}$ &  $0.13^{+0.30}_{-0.56}$ & $13.06^{+0.22}_{-0.29}$ & $43^{+16}_{-9}$ &   $8.99^{+0.24}_{-0.20}$ \\[0.1cm] 
ALESS005.1 & $3.67^{+0.05}_{-0.20}$ & $11.46^{+0.03}_{-0.14}$ & $3.10^{+0.02}_{-0.13}$ & $8.31^{+0.30}_{-0.18}$ & $0.3^{+0.0}_{-0.1}$ & $-0.50^{+0.17}_{-0.17}$ & $12.92^{+0.06}_{-0.05}$ & $40^{+8}_{-3}$ &     $9.08^{+0.24}_{-0.17}$ \\[0.1cm] 
ALESS006.1 & $2.88^{+0.15}_{-0.10}$ & $10.83^{+0.01}_{-0.01}$ & $2.92^{+0.04}_{-0.07}$ & $7.84^{+0.44}_{-0.01}$ & $0.4^{+0.1}_{-0.2}$ & $-0.73^{+0.31}_{-0.02}$ & $12.75^{+0.05}_{-0.03}$ & $43^{+5}_{-1}$ &     $9.17^{+0.02}_{-0.20}$ \\[0.1cm] 
ALESS007.1 & $1.98^{+0.60}_{-0.50}$ & $11.36^{+0.06}_{-0.24}$ & $2.31^{+0.45}_{-0.28}$ & $8.59^{+0.42}_{-0.45}$ & $1.7^{+0.4}_{-0.7}$ & $-0.02^{+0.10}_{-0.30}$ & $12.59^{+0.29}_{-0.28}$ & $35^{+11}_{-7}$ &   $9.16^{+0.12}_{-0.19}$ \\[0.1cm] 
ALESS009.1 & $4.88^{+1.24}_{-1.66}$ & $11.75^{+0.09}_{-0.28}$ & $3.16^{+0.27}_{-0.66}$ & $8.55^{+0.39}_{-0.37}$ & $2.5^{+1.2}_{-0.7}$ &  $0.06^{+0.43}_{-0.34}$ & $13.24^{+0.21}_{-0.46}$ & $53^{+13}_{-19}$ & $8.85^{+0.40}_{-0.13}$ \\[0.1cm] 
ALESS010.1 & $1.42^{+1.41}_{-0.01}$ & $10.08^{+0.01}_{-0.01}$ & $2.28^{+0.53}_{-0.03}$ & $7.71^{+0.53}_{-0.00}$ & $3.2^{+0.0}_{-1.6}$ & $-0.13^{+0.03}_{-0.24}$ & $12.28^{+0.53}_{-0.02}$ & $41^{+5}_{-5}$ &    $9.09^{+0.01}_{-0.14}$ \\[0.1cm] 
ALESS011.1 & $3.03^{+0.25}_{-0.45}$ & $11.46^{+0.10}_{-0.85}$ & $2.73^{+0.23}_{-0.31}$ & $8.73^{+0.26}_{-0.62}$ & $3.3^{+0.7}_{-0.8}$ &  $0.26^{+0.27}_{-0.47}$ & $12.84^{+0.18}_{-0.16}$ & $45^{+3}_{-11}$ &   $9.22^{+0.07}_{-0.24}$ \\[0.1cm] 
ALESS013.1 & $3.22^{+0.75}_{-0.60}$ & $10.92^{+0.11}_{-0.28}$ & $2.62^{+0.20}_{-0.29}$ & $8.37^{+0.44}_{-0.39}$ & $1.8^{+0.3}_{-0.7}$ & $-0.16^{+0.23}_{-0.30}$ & $12.66^{+0.20}_{-0.27}$ & $42^{+8}_{-8}$ &   $9.39^{+0.19}_{-0.22}$ \\[0.1cm] 
ALESS014.1 & $3.38^{+0.24}_{-0.26}$ & $10.96^{+0.13}_{-0.32}$ & $3.08^{+0.13}_{-0.21}$ & $7.98^{+0.65}_{-0.62}$ & $2.7^{+0.4}_{-0.9}$ & $-0.22^{+0.29}_{-0.44}$ & $13.06^{+0.12}_{-0.13}$ & $44^{+14}_{-5}$ & $8.98^{+0.21}_{-0.15}$ \\[0.1cm] 
ALESS015.1 & $2.67^{+0.36}_{-0.69}$ & $11.49^{+0.15}_{-0.39}$ & $2.45^{+0.34}_{-0.59}$ & $8.89^{+0.33}_{-0.50}$ & $3.4^{+1.4}_{-1.1}$ &  $0.38^{+0.41}_{-0.41}$ & $12.64^{+0.23}_{-0.33}$ & $34^{+16}_{-9}$ & $9.39^{+0.18}_{-0.20}$ \\[0.1cm] 
ALESS015.3 & $3.42^{+4.11}_{-2.34}$ & $10.43^{+0.17}_{-0.63}$ & $2.11^{+0.68}_{-1.43}$ & $8.43^{+0.79}_{-0.78}$ & $1.3^{+2.5}_{-1.1}$ & $-0.13^{+0.69}_{-0.59}$ & $12.15^{+0.56}_{-1.20}$ & $42^{+19}_{-16}$ & $8.51^{+0.54}_{-1.00}$ \\[0.1cm] 
ALESS017.1 & $2.12^{+0.35}_{-0.60}$ & $11.37^{+0.01}_{-0.55}$ & $2.46^{+0.21}_{-0.16}$ & $8.79^{+0.20}_{-0.27}$ & $1.6^{+0.1}_{-0.4}$ &  $0.04^{+0.01}_{-0.22}$ & $12.60^{+0.20}_{-0.26}$ & $32^{+4}_{-0}$ & $9.33^{+0.14}_{-0.20}$ \\[0.1cm] 
ALESS018.1 & $2.03^{+0.30}_{-0.20}$ & $11.89^{+0.03}_{-1.29}$ & $2.40^{+0.38}_{-0.14}$ & $9.12^{+0.00}_{-0.50}$ & $2.0^{+0.5}_{-0.7}$ &  $0.25^{+0.00}_{-0.30}$ & $12.69^{+0.21}_{-0.09}$ & $37^{+2}_{-3}$ & $8.99^{+0.15}_{-0.25}$ \\[0.1cm] 
ALESS019.1 & $3.53^{+0.90}_{-0.56}$ & $11.23^{+0.21}_{-0.66}$ & $2.94^{+0.34}_{-0.38}$ & $8.44^{+0.43}_{-0.89}$ & $3.3^{+0.9}_{-1.1}$ &  $0.15^{+0.36}_{-0.68}$ & $13.01^{+0.27}_{-0.28}$ & $50^{+14}_{-12}$ & $8.68^{+0.23}_{-0.12}$ \\[0.1cm] 
ALESS019.2 & $2.17^{+0.36}_{-0.75}$ & $10.40^{+0.17}_{-0.49}$ & $1.81^{+0.37}_{-0.36}$ & $8.68^{+0.41}_{-0.65}$ & $1.0^{+0.8}_{-0.5}$ & $-0.16^{+0.22}_{-0.28}$ & $11.86^{+0.37}_{-0.37}$ & $34^{+16}_{-7}$ & $8.64^{+0.29}_{-0.56}$ \\[0.1cm] 
ALESS022.1 & $2.42^{+0.46}_{-0.75}$ & $11.60^{+0.09}_{-0.26}$ & $2.66^{+0.35}_{-0.55}$ & $8.90^{+0.28}_{-0.45}$ & $2.1^{+0.5}_{-0.5}$ &  $0.11^{+0.22}_{-0.18}$ & $12.77^{+0.31}_{-0.42}$ & $40^{+20}_{-9}$ & $8.85^{+0.25}_{-0.23}$ \\[0.1cm] 
ALESS023.1 & $4.07^{+1.55}_{-0.95}$ & $11.18^{+0.20}_{-0.54}$ & $2.95^{+0.36}_{-0.45}$ & $8.38^{+0.45}_{-0.76}$ & $3.5^{+1.2}_{-1.4}$ &  $0.16^{+0.43}_{-0.64}$ & $13.00^{+0.32}_{-0.34}$ & $49^{+17}_{-15}$ & $8.90^{+0.35}_{-0.25}$ \\[0.1cm] 
ALESS023.7 & $3.08^{+3.85}_{-1.75}$ & $10.28^{+0.22}_{-0.62}$ & $1.96^{+0.70}_{-1.08}$ & $8.45^{+0.72}_{-0.81}$ & $1.3^{+2.5}_{-1.1}$ & $-0.16^{+0.72}_{-0.52}$ & $12.00^{+0.62}_{-0.87}$ & $42^{+18}_{-16}$ & $8.47^{+0.47}_{-0.93}$ \\[0.1cm] 
ALESS025.1 & $2.67^{+0.50}_{-1.25}$ & $10.53^{+0.73}_{-0.09}$ & $2.84^{+0.32}_{-0.50}$ & $8.47^{+0.44}_{-0.56}$ & $2.2^{+0.6}_{-0.7}$ & $-0.11^{+0.24}_{-0.33}$ & $12.96^{+0.24}_{-0.63}$ & $48^{+18}_{-7}$ & $8.96^{+0.21}_{-0.16}$ \\[0.1cm] 
ALESS029.1 & $3.62^{+0.95}_{-0.54}$ & $11.48^{+0.15}_{-0.50}$ & $2.97^{+0.35}_{-0.39}$ & $8.57^{+0.35}_{-0.70}$ & $2.9^{+1.1}_{-1.0}$ &  $0.14^{+0.38}_{-0.55}$ & $13.06^{+0.28}_{-0.21}$ & $49^{+14}_{-11}$ & $8.76^{+0.17}_{-0.14}$ \\[0.1cm] 
ALESS031.1 & $4.22^{+1.46}_{-1.19}$ & $11.45^{+0.12}_{-0.51}$ & $2.92^{+0.39}_{-0.49}$ & $8.56^{+0.33}_{-0.65}$ & $2.8^{+1.0}_{-0.9}$ &  $0.11^{+0.33}_{-0.54}$ & $13.01^{+0.31}_{-0.37}$ & $44^{+18}_{-12}$ & $8.98^{+0.37}_{-0.23}$ \\[0.1cm] 
ALESS035.1 & $3.58^{+0.95}_{-0.86}$ & $11.24^{+0.20}_{-0.56}$ & $2.84^{+0.37}_{-0.54}$ & $8.51^{+0.46}_{-0.82}$ & $2.6^{+1.7}_{-1.5}$ &  $0.07^{+0.54}_{-0.62}$ & $12.92^{+0.30}_{-0.33}$ & $46^{+18}_{-11}$ & $8.64^{+0.24}_{-0.12}$ \\[0.1cm] 
ALESS035.2 & $4.57^{+3.00}_{-2.15}$ & $10.21^{+0.23}_{-0.55}$ & $2.12^{+0.51}_{-0.61}$ & $8.26^{+0.62}_{-0.74}$ & $4.9^{+4.1}_{-3.5}$ &  $0.42^{+0.87}_{-0.82}$ & $12.17^{+0.47}_{-0.48}$ & $43^{+19}_{-15}$ & $8.40^{+0.55}_{-0.42}$ \\[0.1cm] 
ALESS037.1 & $2.72^{+0.36}_{-0.25}$ & $11.30^{+0.07}_{-0.30}$ & $2.80^{+0.31}_{-0.61}$ & $8.51^{+0.29}_{-0.45}$ & $1.8^{+0.4}_{-0.5}$ & $-0.12^{+0.14}_{-0.34}$ & $12.84^{+0.26}_{-0.29}$ & $53^{+14}_{-16}$ & $8.49^{+0.25}_{-0.15}$ \\[0.1cm] 
ALESS037.2 & $3.83^{+0.64}_{-0.55}$ & $10.55^{+0.10}_{-0.27}$ & $2.33^{+0.52}_{-0.44}$ & $8.27^{+0.47}_{-0.64}$ & $0.8^{+0.8}_{-0.5}$ & $-0.39^{+0.25}_{-0.30}$ & $12.31^{+0.50}_{-0.45}$ & $47^{+19}_{-16}$ & $8.34^{+0.52}_{-0.47}$ \\[0.1cm] 
ALESS039.1 & $2.33^{+0.50}_{-0.35}$ & $10.65^{+0.09}_{-0.19}$ & $2.34^{+0.36}_{-0.20}$ & $8.33^{+0.44}_{-0.35}$ & $1.4^{+0.3}_{-0.4}$ & $-0.28^{+0.18}_{-0.30}$ & $12.31^{+0.35}_{-0.13}$ & $32^{+16}_{-4}$ & $9.03^{+0.13}_{-0.13}$ \\[0.1cm] 
ALESS041.1 & $2.17^{+0.61}_{-0.65}$ & $11.35^{+0.20}_{-0.52}$ & $2.26^{+0.62}_{-0.49}$ & $8.91^{+0.38}_{-0.68}$ & $1.7^{+1.2}_{-1.2}$ &  $0.08^{+0.39}_{-0.55}$ & $12.43^{+0.47}_{-0.27}$ & $34^{+23}_{-7}$ & $9.06^{+0.26}_{-0.26}$ \\[0.1cm] 
ALESS041.3 & $2.97^{+4.46}_{-2.40}$ & $9.86^{+0.51}_{-0.79}$ &   $2.13^{+0.84}_{-2.70}$ & $8.35^{+1.13}_{-1.08}$ & $0.7^{+2.8}_{-0.6}$ & $-0.24^{+0.80}_{-0.60}$ & $12.11^{+0.71}_{-2.46}$ & $42^{+19}_{-16}$ & $8.42^{+0.73}_{-2.42}$ \\[0.1cm] 
ALESS043.1 & $2.08^{+0.50}_{-0.60}$ & $11.31^{+0.09}_{-0.20}$ & $1.90^{+0.56}_{-0.39}$ & $9.07^{+0.28}_{-0.46}$ & $2.8^{+1.0}_{-0.8}$ &  $0.33^{+0.32}_{-0.28}$ & $12.15^{+0.40}_{-0.21}$ & $34^{+19}_{-7}$ & $8.72^{+0.26}_{-0.30}$ \\[0.1cm] 
ALESS045.1 & $3.17^{+0.86}_{-0.59}$ & $11.71^{+0.11}_{-0.26}$ & $2.51^{+0.39}_{-0.46}$ & $8.85^{+0.30}_{-0.41}$ & $2.8^{+0.7}_{-0.7}$ &  $0.28^{+0.29}_{-0.26}$ & $12.77^{+0.26}_{-0.30}$ & $38^{+15}_{-9}$ & $8.93^{+0.22}_{-0.22}$ \\[0.1cm] 
ALESS049.1 & $2.83^{+0.14}_{-0.05}$ & $10.58^{+0.12}_{-0.22}$ & $2.83^{+0.09}_{-0.05}$ & $7.75^{+0.83}_{-0.19}$ & $1.9^{+0.3}_{-1.1}$ & $-0.33^{+0.10}_{-0.31}$ & $12.83^{+0.04}_{-0.07}$ & $46^{+0}_{-3}$ & $9.04^{+0.04}_{-0.12}$ \\[0.1cm] 
ALESS049.2 & $2.67^{+0.66}_{-0.25}$ & $11.02^{+0.15}_{-0.36}$ & $2.54^{+0.39}_{-0.25}$ & $8.60^{+0.52}_{-0.84}$ & $1.6^{+0.3}_{-0.6}$ & $-0.11^{+0.18}_{-0.42}$ & $12.60^{+0.32}_{-0.19}$ & $51^{+13}_{-8}$ & $8.28^{+0.21}_{-0.21}$ \\[0.1cm] 
ALESS051.1 & $1.33^{+0.19}_{-0.10}$ & $11.17^{+0.07}_{-0.24}$ & $1.97^{+0.20}_{-0.33}$ & $9.05^{+0.27}_{-0.36}$ & $1.6^{+0.4}_{-0.6}$ &  $0.11^{+0.17}_{-0.19}$ & $12.13^{+0.15}_{-0.12}$ & $31^{+12}_{-5}$ & $9.16^{+0.09}_{-0.13}$ \\[0.1cm] 
ALESS055.1 & $2.28^{+0.25}_{-0.20}$ & $9.97^{+0.27}_{-0.21}$ &   $2.22^{+0.18}_{-0.22}$ & $8.00^{+0.87}_{-0.70}$ & $1.1^{+0.8}_{-1.0}$ & $-0.39^{+0.16}_{-0.33}$ & $12.15^{+0.21}_{-0.13}$ & $29^{+15}_{-2}$ & $9.05^{+0.15}_{-0.14}$ \\[0.1cm] 
ALESS055.2 & $4.68^{+2.89}_{-1.96}$ & $10.42^{+0.22}_{-0.55}$ & $2.36^{+0.48}_{-0.58}$ & $8.24^{+0.59}_{-0.73}$ & $5.4^{+3.7}_{-3.4}$ &  $0.51^{+0.79}_{-0.80}$ & $12.40^{+0.43}_{-0.47}$ & $44^{+18}_{-16}$ & $8.61^{+0.56}_{-0.42}$ \\[0.1cm] 
ALESS055.5 & $2.33^{+0.14}_{-1.06}$ & $10.15^{+0.04}_{-0.88}$ & $0.80^{+1.00}_{-0.20}$ & $9.13^{+0.41}_{-1.60}$ & $0.0^{+2.0}_{-0.0}$ & $-0.15^{+0.07}_{-0.56}$ & $10.19^{+1.53}_{-0.81}$ & $44^{+23}_{-15}$ & $6.38^{+2.11}_{-0.38}$ \\[0.1cm] 
ALESS057.1 & $1.98^{+0.60}_{-0.50}$ & $9.92^{+0.27}_{-0.13}$ &   $2.39^{+0.41}_{-0.18}$ & $7.70^{+1.09}_{-0.27}$ & $2.4^{+0.6}_{-1.4}$ & $-0.48^{+0.40}_{-0.30}$ & $12.37^{+0.40}_{-0.20}$ & $42^{+15}_{-14}$ & $9.00^{+0.12}_{-0.42}$ \\[0.1cm] 
ALESS059.2 & $1.48^{+0.25}_{-0.21}$ & $9.76^{+0.10}_{-0.24}$ &   $1.87^{+0.12}_{-0.23}$ & $7.97^{+0.76}_{-0.30}$ & $2.5^{+0.5}_{-1.2}$ & $-0.16^{+0.10}_{-0.32}$ & $11.88^{+0.10}_{-0.22}$ & $34^{+10}_{-7}$ & $8.73^{+0.16}_{-0.22}$ \\[0.1cm] 
ALESS061.1 & $6.12^{+0.26}_{-1.44}$ & $10.58^{+0.02}_{-0.25}$ & $3.14^{+0.14}_{-0.30}$ & $7.27^{+0.89}_{-0.00}$ & $0.5^{+0.2}_{-0.3}$ & $-0.29^{+0.00}_{-0.71}$ & $13.00^{+0.10}_{-0.32}$ & $57^{+9}_{-14}$ & $8.57^{+0.62}_{-0.22}$ \\[0.1cm] 
ALESS063.1 & $2.08^{+0.20}_{-0.46}$ & $11.20^{+0.01}_{-0.19}$ & $1.98^{+0.00}_{-0.34}$ & $9.01^{+0.22}_{-0.63}$ & $1.2^{+0.6}_{-0.0}$ &  $0.03^{+0.26}_{-0.14}$ & $12.14^{+0.08}_{-0.19}$ & $37^{+0}_{-13}$ & $9.26^{+0.13}_{-0.07}$ \\[0.1cm] 
ALESS065.1 & $5.68^{+1.79}_{-2.76}$ & $10.74^{+0.17}_{-0.48}$ & $2.64^{+0.44}_{-0.58}$ & $8.23^{+0.47}_{-0.65}$ & $1.9^{+1.8}_{-1.1}$ & $-0.17^{+0.54}_{-0.49}$ & $12.66^{+0.40}_{-0.47}$ & $44^{+17}_{-16}$ & $8.91^{+0.56}_{-0.48}$ \\[0.1cm] 
ALESS066.1 & $1.98^{+0.49}_{-1.00}$ & $10.07^{+0.35}_{-0.27}$ & $2.61^{+0.45}_{-0.50}$ & $7.75^{+0.97}_{-0.34}$ & $0.5^{+0.3}_{-0.3}$ & $-0.80^{+0.48}_{-0.20}$ & $12.44^{+0.34}_{-0.45}$ & $49^{+10}_{-10}$ & $8.71^{+0.12}_{-0.27}$ \\[0.1cm] 
ALESS067.1 & $2.08^{+0.30}_{-0.35}$ & $11.38^{+0.30}_{-0.84}$ & $2.57^{+0.21}_{-0.11}$ & $8.91^{+0.20}_{-0.91}$ & $1.7^{+0.3}_{-0.7}$ &  $0.02^{+0.11}_{-0.46}$ & $12.70^{+0.13}_{-0.24}$ & $38^{+5}_{-4}$ & $8.87^{+0.19}_{-0.13}$ \\[0.1cm] 
ALESS067.2 & $1.52^{+1.06}_{-0.25}$ & $9.81^{+0.087}_{-0.25}$ & $1.85^{+0.35}_{-0.40}$ & $8.01^{+0.94}_{-0.37}$ & $2.0^{+0.5}_{-1.1}$ & $-0.44^{+0.38}_{-0.14}$ & $11.85^{+0.36}_{-0.28}$ & $42^{+10}_{-16}$ & $8.62^{+0.25}_{-0.64}$ \\[0.1cm] 
ALESS068.1 & $3.78^{+1.90}_{-1.06}$ & $10.97^{+0.19}_{-0.57}$ & $2.61^{+0.44}_{-0.53}$ & $8.48^{+0.44}_{-0.75}$ & $3.2^{+1.4}_{-1.2}$ &  $0.16^{+0.47}_{-0.64}$ & $12.69^{+0.37}_{-0.40}$ & $47^{+17}_{-15}$ & $8.67^{+0.40}_{-0.34}$ \\[0.1cm]
\enddata
\end{deluxetable*}
\end{small}

\setcounter{table}{0}
\begin{deluxetable*}{lrrrrrrrrr}
\tablecolumns{5}
\tablewidth{0pt}
\tablecaption{(continued)}
\tablehead{
\colhead{ID} &
\colhead{$z_\mathrm{phot}$} &
\colhead{$\log(M_\ast/M_\sun)$} &
\colhead{$\log(\mathrm{SFR}/M_\sun \mathrm{yr}^{-1})$} &
\colhead{$\log(\mathrm{age}_M/\mathrm{yr})$} &
\colhead{$A_V$} &
\colhead{$\log(M_\ast/L_\mathrm{H})$} &
\colhead{$\log(L_\mathrm{dust}/L_\sun)$} &
\colhead{$T_\mathrm{dust}/K$} &
\colhead{$\log(M_\mathrm{dust}/M_\sun)$}
\\
\\
\colhead{(1)} &
\colhead{(2)} &
\colhead{(3)} &
\colhead{(4)} &
\colhead{(5)} &
\colhead{(6)} &
\colhead{(7)} &
\colhead{(8)} &
\colhead{(9)} &
\colhead{(10)}
}
\startdata
ALESS069.1 & $2.83^{+0.50}_{-0.80}$ & $11.43^{+0.11}_{-0.29}$ & $2.29^{+0.58}_{-0.70}$ & $8.87^{+0.36}_{-0.45}$ & $2.7^{+1.1}_{-0.8}$ & $0.26^{+0.36}_{-0.28}$ & $12.50^{+0.47}_{-0.50}$ & $36^{+23}_{-10}$ & $8.99^{+0.33}_{-0.36}$ \\[0.1cm] 
ALESS069.2 & $4.38^{+3.15}_{-1.76}$ & $10.38^{+0.22}_{-0.52}$ & $2.32^{+0.51}_{-0.57}$ & $8.23^{+0.59}_{-0.72}$ & $4.2^{+4.7}_{-3.1}$ & $0.28^{+0.96}_{-0.78}$ & $12.37^{+0.46}_{-0.47}$ & $43^{+18}_{-15}$ & $8.64^{+0.52}_{-0.44}$ \\[0.1cm] 
ALESS069.3 & $4.68^{+2.89}_{-2.06}$ & $10.36^{+0.22}_{-0.55}$ & $2.29^{+0.50}_{-0.59}$ & $8.24^{+0.60}_{-0.73}$ & $5.3^{+3.8}_{-3.4}$ & $0.49^{+0.81}_{-0.80}$ & $12.34^{+0.45}_{-0.48}$ & $44^{+18}_{-16}$ & $8.55^{+0.56}_{-0.41}$ \\[0.1cm] 
ALESS070.1 & $1.58^{+0.64}_{-0.20}$ & $10.63^{+0.00}_{-0.22}$ & $2.65^{+0.14}_{-0.17}$ & $8.01^{+0.89}_{-0.29}$ & $2.5^{+0.0}_{-1.5}$ & $-0.47^{+0.40}_{-0.00}$ & $12.63^{+0.21}_{-0.17}$ & $46^{+0}_{-4}$ & $8.84^{+0.33}_{-0.00}$ \\[0.1cm] 
ALESS071.1 & $1.77^{+0.21}_{-0.39}$ & $11.39^{+0.18}_{-0.35}$ & $2.75^{+0.14}_{-0.30}$ & $8.80^{+0.32}_{-0.60}$ & $2.5^{+0.4}_{-0.5}$ & $0.11^{+0.17}_{-0.37}$ & $12.82^{+0.10}_{-0.27}$ & $43^{+3}_{-9}$ & $8.71^{+0.19}_{-0.07}$ \\[0.1cm] 
ALESS071.3 & $2.28^{+0.39}_{-0.80}$ & $9.64^{+0.132}_{-0.42}$ & $1.62^{+0.52}_{-0.95}$ & $8.18^{+0.78}_{-0.83}$ & $0.8^{+0.7}_{-0.7}$ & $-0.37^{+0.30}_{-0.35}$ & $11.61^{+0.45}_{-1.15}$ & $37^{+23}_{-10}$ & $8.27^{+0.45}_{-1.87}$ \\[0.1cm] 
ALESS072.1 & $5.82^{+1.65}_{-2.79}$ & $10.95^{+0.14}_{-0.41}$ & $2.74^{+0.41}_{-0.61}$ & $8.31^{+0.43}_{-0.58}$ & $1.7^{+1.6}_{-0.9}$ & $-0.2^{+0.54}_{-0.43}$ & $12.77^{+0.38}_{-0.48}$ & $44^{+18}_{-14}$ & $8.93^{+0.61}_{-0.44}$ \\[0.1cm] 
ALESS073.1 & $4.78^{+0.40}_{-0.50}$ & $10.64^{+0.05}_{-0.22}$ & $2.90^{+0.30}_{-0.36}$ & $7.67^{+0.79}_{-0.31}$ & $1.5^{+0.5}_{-0.7}$ & $-0.41^{+0.19}_{-0.26}$ & $12.89^{+0.28}_{-0.35}$ & $46^{+17}_{-14}$ & $8.97^{+0.34}_{-0.32}$ \\[0.1cm]
ALESS074.1 & $1.62^{+0.85}_{-0.24}$ & $10.68^{+0.28}_{-0.72}$ & $2.17^{+0.45}_{-0.22}$ & $8.87^{+0.45}_{-1.07}$ & $2.6^{+0.6}_{-1.1}$ & $0.12^{+0.28}_{-0.74}$ & $12.17^{+0.49}_{-0.11}$ & $32^{+12}_{-5}$ & $9.04^{+0.20}_{-0.27}$ \\[0.1cm] 
ALESS075.1 & $1.98^{+0.24}_{-0.10}$ & $10.48^{+0.00}_{-0.21}$ & $2.65^{+0.13}_{-0.14}$ & $7.74^{+0.56}_{-0.19}$ & $1.6^{+0.5}_{-0.5}$ & $-0.68^{+0.44}_{-0.00}$ & $12.58^{+0.17}_{-0.11}$ & $39^{+8}_{-1}$ & $8.76^{+0.23}_{-0.07}$ \\[0.1cm] 
ALESS075.4 & $2.28^{+0.60}_{-1.11}$ & $9.28^{+0.22}_{-0.73}$ & $1.57^{+0.56}_{-1.10}$ & $8.01^{+0.81}_{-0.62}$ & $0.5^{+1.0}_{-0.4}$ & $-0.44^{+0.31}_{-0.40}$ & $11.52^{+0.51}_{-1.27}$ & $38^{+25}_{-10}$ & $8.11^{+0.59}_{-2.06}$ \\[0.1cm] 
ALESS076.1 & $3.97^{+1.71}_{-0.94}$ & $11.01^{+0.20}_{-0.49}$ & $2.84^{+0.37}_{-0.46}$ & $8.31^{+0.49}_{-0.66}$ & $3.5^{+1.4}_{-1.7}$ & $0.17^{+0.47}_{-0.53}$ & $12.91^{+0.32}_{-0.42}$ & $44^{+17}_{-13}$ & $8.90^{+0.40}_{-0.25}$ \\[0.1cm] 
ALESS079.1 & $3.53^{+1.09}_{-0.86}$ & $11.56^{+0.12}_{-0.30}$ & $2.58^{+0.44}_{-0.54}$ & $8.76^{+0.30}_{-0.35}$ & $2.7^{+0.8}_{-0.9}$ & $0.20^{+0.30}_{-0.32}$ & $12.78^{+0.30}_{-0.36}$ & $43^{+17}_{-11}$ & $8.66^{+0.28}_{-0.18}$ \\[0.1cm] 
ALESS079.2 & $1.88^{+0.01}_{-0.00}$ & $11.71^{+0.00}_{-0.37}$ & $2.38^{+0.00}_{-0.01}$ & $9.17^{+0.00}_{-0.03}$ & $1.4^{+0.3}_{-0.0}$ & $0.12^{+0.01}_{-0.00}$ & $12.50^{+0.00}_{-0.01}$ & $43^{+2}_{-1}$ & $8.35^{+0.12}_{-0.17}$ \\[0.1cm] 
ALESS079.4 & $1.83^{+5.24}_{-1.21}$ & $8.38^{+0.61}_{-0.38}$ & $0.60^{+1.95}_{-1.82}$ & $8.32^{+0.81}_{-0.81}$ & $1.0^{+7.3}_{-1.0}$ & $-0.22^{+1.32}_{-0.50}$ & $10.58^{+2.00}_{-2.24}$ & $43^{+21}_{-15}$ & $6.94^{+1.96}_{-0.94}$ \\[0.1cm] 
ALESS080.1 & $2.58^{+1.14}_{-0.41}$ & $11.12^{+0.04}_{-0.75}$ & $2.43^{+0.56}_{-0.31}$ & $8.71^{+0.38}_{-0.64}$ & $2.0^{+0.4}_{-0.6}$ & $0.04^{+0.14}_{-0.39}$ & $12.50^{+0.52}_{-0.25}$ & $36^{+24}_{-2}$ & $8.73^{+0.29}_{-0.29}$ \\[0.1cm] 
ALESS080.2 & $1.48^{+0.50}_{-0.10}$ & $10.01^{+0.05}_{-0.04}$ & $2.19^{+0.08}_{-0.16}$ & $8.08^{+1.01}_{-0.36}$ & $3.0^{+0.3}_{-0.9}$ & $-0.43^{+0.46}_{-0.07}$ & $12.17^{+0.07}_{-0.16}$ & $31^{+25}_{-2}$ & $8.92^{+0.23}_{-0.85}$ \\[0.1cm] 
ALESS082.1 & $3.47^{+2.65}_{-1.95}$ & $10.98^{+0.17}_{-0.46}$ & $2.39^{+0.53}_{-1.17}$ & $8.63^{+0.55}_{-0.62}$ & $1.9^{+1.8}_{-1.3}$ & $-0.05^{+0.59}_{-0.42}$ & $12.46^{+0.46}_{-0.88}$ & $47^{+20}_{-20}$ & $8.36^{+0.54}_{-0.35}$ \\[0.1cm] 
ALESS083.4 & $2.72^{+4.06}_{-1.74}$ & $10.75^{+0.19}_{-0.55}$ & $2.09^{+1.02}_{-1.30}$ & $8.68^{+0.71}_{-0.83}$ & $0.4^{+2.2}_{-0.4}$ & $-0.18^{+0.65}_{-0.53}$ & $12.04^{+0.68}_{-0.88}$ & $41^{+25}_{-16}$ & $8.34^{+0.43}_{-0.54}$ \\[0.1cm] 
ALESS084.1 & $1.48^{+0.10}_{-0.21}$ & $10.01^{+0.15}_{-0.09}$ & $2.22^{+0.09}_{-0.06}$ & $7.79^{+0.22}_{-0.12}$ & $2.4^{+0.3}_{-0.1}$ & $-0.48^{+0.07}_{-0.03}$ & $12.19^{+0.08}_{-0.07}$ & $42^{+2}_{-11}$ & $9.11^{+0.11}_{-0.26}$ \\[0.1cm] 
ALESS084.2 & $1.73^{+0.44}_{-0.15}$ & $10.77^{+0.21}_{-0.18}$ & $1.91^{+0.20}_{-0.21}$ & $8.89^{+0.39}_{-0.18}$ & $1.8^{+0.2}_{-0.8}$ & $-0.04^{+0.26}_{-0.05}$ & $11.98^{+0.21}_{-0.19}$ & $31^{+13}_{-4}$ & $8.89^{+0.23}_{-0.57}$ \\[0.1cm] 
ALESS087.1 & $1.38^{+0.45}_{-0.30}$ & $10.29^{+0.14}_{-0.45}$ & $2.34^{+0.37}_{-0.33}$ & $8.03^{+0.69}_{-0.54}$ & $2.3^{+1.0}_{-0.9}$ & $-0.21^{+0.29}_{-0.53}$ & $12.36^{+0.27}_{-0.31}$ & $58^{+3}_{-18}$ & $8.35^{+0.16}_{-0.40}$ \\[0.1cm] 
ALESS087.3 & $4.68^{+2.89}_{-1.90}$ & $10.43^{+0.22}_{-0.54}$ & $2.37^{+0.49}_{-0.56}$ & $8.23^{+0.59}_{-0.72}$ & $5.4^{+3.7}_{-3.4}$ & $0.51^{+0.79}_{-0.80}$ & $12.42^{+0.43}_{-0.47}$ & $44^{+18}_{-16}$ & $8.63^{+0.56}_{-0.42}$ \\[0.1cm] 
ALESS088.1 & $1.58^{+0.15}_{-0.06}$ & $10.06^{+0.17}_{-0.00}$ & $2.34^{+0.23}_{-0.08}$ & $8.01^{+0.00}_{-0.39}$ & $2.5^{+0.0}_{-0.4}$ & $-0.47^{+0.00}_{-0.31}$ & $12.31^{+0.13}_{-0.07}$ & $49^{+1}_{-16}$ & $9.13^{+0.01}_{-0.18}$ \\[0.1cm] 
ALESS088.2 & $4.28^{+3.10}_{-1.80}$ & $10.64^{+0.14}_{-0.40}$ &  $2.18^{+0.46}_{-0.61}$ & $8.51^{+0.48}_{-0.25}$ & $6.6^{+2.6}_{-4.0}$ & $0.83^{+0.61}_{-0.80}$ & $12.29^{+0.42}_{-0.50}$ & $40^{+20}_{-13}$ & $8.58^{+0.53}_{-0.40}$ \\[0.1cm] 
ALESS088.5 & $2.47^{+0.61}_{-0.64}$ & $10.89^{+0.13}_{-0.45}$ & $2.65^{+0.27}_{-0.46}$ & $8.32^{+0.60}_{-0.55}$ & $2.2^{+0.5}_{-0.8}$ & $-0.10^{+0.24}_{-0.34}$ & $12.68^{+0.25}_{-0.43}$ & $46^{+12}_{-9}$ & $8.52^{+0.38}_{-0.20}$ \\[0.1cm] 
ALESS088.11 & $2.42^{+0.25}_{-0.2}$ & $10.42^{+0.05}_{-0.31}$ & $2.46^{+0.17}_{-0.19}$ & $7.99^{+0.62}_{-0.62}$ & $1.2^{+0.4}_{-0.6}$ & $-0.48^{+0.27}_{-0.25}$ & $12.44^{+0.11}_{-0.18}$ & $38^{+6}_{-4}$ & $8.76^{+0.14}_{-0.25}$ \\[0.1cm] 
ALESS092.2 & $1.58^{+0.84}_{-0.85}$ & $9.16^{+0.26}_{-0.63}$ & $1.27^{+0.84}_{-1.18}$ & $8.13^{+0.79}_{-0.63}$ & $0.6^{+0.9}_{-0.4}$ & $-0.41^{+0.34}_{-0.37}$ & $11.2^{+0.89}_{-1.15}$ & $40^{+24}_{-13}$ & $7.71^{+1.16}_{-1.71}$ \\[0.1cm] 
ALESS094.1 & $2.38^{+0.24}_{-0.55}$ & $10.29^{+0.06}_{-0.34}$ & $2.40^{+0.21}_{-0.25}$ & $8.01^{+0.91}_{-0.40}$ & $2.5^{+0.1}_{-0.7}$ & $-0.47^{+0.23}_{-0.31}$ & $12.38^{+0.18}_{-0.26}$ & $43^{+16}_{-11}$ & $8.92^{+0.22}_{-0.33}$ \\[0.1cm] 
ALESS098.1 & $3.33^{+0.01}_{-2.00}$ & $11.87^{+0.12}_{-0.21}$ & $3.40^{+0.01}_{-1.28}$ & $8.56^{+0.66}_{-0.15}$ & $2.3^{+2.1}_{-0.2}$ & $-0.07^{+0.79}_{-0.14}$ & $13.43^{+0.02}_{-0.96}$ & $62^{+0}_{-30}$ & $8.72^{+0.32}_{-0.03}$ \\[0.1cm] 
ALESS099.1 & $4.62^{+2.95}_{-2.00}$ & $10.36^{+0.23}_{-0.55}$ & $2.30^{+0.49}_{-0.57}$ & $8.24^{+0.59}_{-0.73}$ & $5.4^{+3.7}_{-3.5}$ & $0.50^{+0.80}_{-0.81}$ & $12.35^{+0.45}_{-0.47}$ & $44^{+18}_{-16}$ & $8.55^{+0.56}_{-0.40}$ \\[0.1cm] 
ALESS102.1 & $2.42^{+0.30}_{-0.44}$ & $11.61^{+0.07}_{-0.14}$ & $2.15^{+0.40}_{-0.42}$ & $9.04^{+0.26}_{-0.29}$ & $1.6^{+0.5}_{-0.4}$ & $0.12^{+0.23}_{-0.13}$ & $12.35^{+0.32}_{-0.23}$ & $37^{+21}_{-8}$ & $8.75^{+0.27}_{-0.21}$ \\[0.1cm] 
ALESS103.3 & $4.57^{+3.00}_{-2.15}$ & $10.21^{+0.22}_{-0.57}$ & $2.12^{+0.52}_{-0.62}$ & $8.26^{+0.61}_{-0.74}$ & $5.1^{+4.0}_{-3.5}$ & $0.45^{+0.84}_{-0.81}$ & $12.17^{+0.47}_{-0.49}$ & $43^{+19}_{-15}$ & $8.40^{+0.56}_{-0.42}$ \\[0.1cm] 
ALESS107.1 & $3.42^{+0.36}_{-0.45}$ & $11.00^{+0.09}_{-0.39}$ & $2.73^{+0.29}_{-0.37}$ & $8.31^{+0.49}_{-0.84}$ & $0.7^{+0.4}_{-0.3}$ & $-0.42^{+0.27}_{-0.38}$ & $12.69^{+0.24}_{-0.28}$ & $55^{+16}_{-15}$ & $8.28^{+0.32}_{-0.21}$ \\[0.1cm] 
ALESS107.3 & $2.22^{+1.06}_{-1.24}$ & $9.33^{+0.23}_{-0.83}$ & $1.59^{+0.66}_{-1.46}$ & $8.03^{+0.84}_{-0.64}$ & $0.7^{+1.0}_{-0.6}$ & $-0.36^{+0.30}_{-0.37}$ & $11.57^{+0.61}_{-1.55}$ & $38^{+23}_{-11}$ & $8.24^{+0.51}_{-2.24}$ \\[0.1cm] 
ALESS110.1 & $3.58^{+0.74}_{-0.46}$ & $11.05^{+0.24}_{-0.51}$ & $2.95^{+0.18}_{-0.26}$ & $8.29^{+0.54}_{-0.9.0}$ & $3.5^{+0.9}_{-0.9}$ & $0.09^{+0.40}_{-0.68}$ & $12.98^{+0.13}_{-0.20}$ & $66^{+1}_{-15}$ & $8.71^{+0.35}_{-0.26}$ \\[0.1cm] 
ALESS110.5 & $3.62^{+3.70}_{-2.74}$ & $10.55^{+0.16}_{-0.65}$ & $2.18^{+0.60}_{-1.58}$ & $8.45^{+0.85}_{-0.63}$ & $1.4^{+3.4}_{-1.0}$ & $-0.10^{+0.92}_{-0.51}$ & $12.25^{+0.52}_{-1.33}$ & $42^{+18}_{-16}$ & $8.61^{+0.55}_{-0.95}$ \\[0.1cm] 
ALESS112.1 & $2.72^{+0.25}_{-1.14}$ & $11.01^{+0.15}_{-0.31}$ & $2.71^{+0.24}_{-0.85}$ & $8.45^{+0.67}_{-0.71}$ & $1.0^{+0.9}_{-0.6}$ & $-0.31^{+0.63}_{-0.5}$ & $12.70^{+0.14}_{-0.57}$ & $36^{+13}_{-10}$ & $9.28^{+0.20}_{-0.19}$ \\[0.1cm] 
ALESS114.1 & $3.42^{+0.96}_{-1.75}$ & $10.92^{+0.18}_{-0.51}$ & $2.76^{+0.39}_{-0.94}$ & $8.28^{+0.70}_{-0.73}$ & $4.2^{+2.4}_{-2.1}$ & $0.33^{+0.72}_{-0.61}$ & $12.84^{+0.33}_{-0.78}$ & $48^{+16}_{-14}$ & $8.55^{+0.39}_{-0.18}$ \\[0.1cm] 
ALESS114.2 & $1.58^{+0.59}_{-0.16}$ & $11.32^{+0.14}_{-0.27}$ & $2.38^{+0.27}_{-0.16}$ & $8.97^{+0.20}_{-0.39}$ & $1.6^{+0.3}_{-0.6}$ & $0.04^{+0.10}_{-0.22}$ & $12.48^{+0.32}_{-0.09}$ & $40^{+5}_{-6}$ & $8.66^{+0.13}_{-0.26}$ \\[0.1cm] 
ALESS115.1 & $3.83^{+1.24}_{-0.75}$ & $11.34^{+0.20}_{-0.56}$ & $2.99^{+0.39}_{-0.48}$ & $8.48^{+0.43}_{-0.80}$ & $2.9^{+1.4}_{-1.4}$ & $0.12^{+0.44}_{-0.65}$ & $13.07^{+0.31}_{-0.27}$ & $46^{+18}_{-10}$ & $8.82^{+0.30}_{-0.14}$ \\[0.1cm] 
ALESS116.1 & $3.58^{+0.80}_{-0.80}$ & $10.92^{+0.17}_{-0.44}$ & $2.74^{+0.34}_{-0.40}$ & $8.28^{+0.55}_{-0.65}$ & $3.5^{+1.3}_{-1.5}$ & $0.15^{+0.47}_{-0.52}$ & $12.80^{+0.28}_{-0.33}$ & $48^{+16}_{-12}$ & $8.52^{+0.27}_{-0.17}$ \\[0.1cm] 
ALESS116.2 & $3.58^{+0.70}_{-0.75}$ & $11.19^{+0.16}_{-0.50}$ & $2.67^{+0.28}_{-0.43}$ & $8.58^{+0.32}_{-0.64}$ & $3.2^{+1.1}_{-0.9}$ & $0.22^{+0.31}_{-0.50}$ & $12.77^{+0.21}_{-0.28}$ & $44^{+16}_{-10}$ & $8.60^{+0.30}_{-0.16}$ \\[0.1cm] 
ALESS118.1 & $3.47^{+1.41}_{-1.49}$ & $10.76^{+0.15}_{-0.33}$ & $2.93^{+0.33}_{-0.82}$ & $7.93^{+0.64}_{-0.51}$ & $1.8^{+0.6}_{-0.7}$ & $-0.43^{+0.38}_{-0.35}$ & $12.89^{+0.34}_{-0.72}$ & $56^{+13}_{-20}$ & $8.53^{+0.37}_{-0.22}$ \\[0.1cm] 
ALESS119.1 & $3.47^{+0.91}_{-0.50}$ & $10.41^{+0.10}_{-0.21}$ & $2.73^{+0.31}_{-0.31}$ & $7.69^{+0.79}_{-0.33}$ & $2.0^{+0.6}_{-1.1}$ & $-0.46^{+0.32}_{-0.22}$ & $12.69^{+0.31}_{-0.23}$ & $35^{+19}_{-7}$ & $9.34^{+0.25}_{-0.26}$ \\[0.1cm] 
ALESS122.1 & $2.03^{+0.19}_{-0.45}$ & $10.89^{+0.21}_{-0.21}$ & $2.84^{+0.17}_{-0.16}$ & $8.19^{+0.79}_{-0.80}$ & $1.7^{+0.3}_{-0.9}$ & $-0.15^{+0.20}_{-0.63}$ & $12.92^{+0.13}_{-0.25}$ & $41^{+6}_{-2}$ & $8.71^{+0.15}_{-0.10}$ \\[0.1cm] 
ALESS124.1 & $2.42^{+0.80}_{-0.94}$ & $11.13^{+0.16}_{-0.52}$ & $2.14^{+0.54}_{-0.69}$ & $8.86^{+0.43}_{-0.60}$ & $3.4^{+1.6}_{-0.9}$ & $0.36^{+0.48}_{-0.51}$ & $12.35^{+0.40}_{-0.49}$ & $34^{+15}_{-8}$ & $8.91^{+0.31}_{-0.31}$ \\[0.1cm] 
ALESS124.4 & $3.42^{+2.01}_{-1.00}$ & $10.43^{+0.27}_{-0.77}$ & $2.48^{+0.47}_{-0.52}$ & $8.10^{+0.74}_{-0.67}$ & $5.0^{+4.1}_{-4.9}$ & $0.37^{+0.96}_{-1.12}$ & $12.51^{+0.41}_{-0.49}$ & $44^{+22}_{-12}$ & $8.45^{+0.41}_{-0.35}$ \\[0.1cm] 
ALESS126.1 & $1.88^{+0.65}_{-0.50}$ & $10.07^{+0.80}_{-0.08}$ & $2.11^{+0.29}_{-0.22}$ & $8.65^{+0.60}_{-0.94}$ & $2.2^{+1.1}_{-1.7}$ & $-0.06^{+0.29}_{-0.51}$ & $12.17^{+0.28}_{-0.21}$ & $59^{+6}_{-27}$ & $8.44^{+0.52}_{-0.49}$ \\[0.1cm] 
\enddata
\tablecomments{(1) Catalog ID; (2) photometric redshift; (3) stellar mass; (4) star formation rate; (5) mass-weighted age; (6) average $V$-band dust attenuation; (7) $H$-band mass-to-light ratio; (8) total dust luminosity; (9) luminosity-averaged dust temperature; (10) total dust mass.}
\end{deluxetable*}

\end{document}